\documentclass{elsart}
\usepackage{graphicx,natbib,amssymb}

\def\figdir{.}

\def\url#1{{\ttfamily\def\/{/\discretionary{}{}{}}#1}}

\begin{document}

\begin{frontmatter}

\title{Multi--dimensional Cosmological Radiative Transfer with  a
Variable Eddington Tensor Formalism}

\author{Nickolay Y.\ Gnedin\thanksref{email1}}
\address{Center for Astrophysics and Space Astronomy, 
University of Colorado, Boulder, CO 80309}
\thanks[email1]{E-mail: ; {\tt gnedin@casa.colorado.edu}}

\author{Tom Abel\thanksref{email2}}
\address{Harvard Smithsonian Center for Astrophysics, Cambridge, 
MA, US 02138}
\thanks[email2]{E-mail: ; {\tt RT@TomAbel.com}}

\begin{abstract}
We present a new approach to numerically model continuum radiative
transfer based on the Optically Thin Variable Eddington Tensor
(OTVET) approximation. Our method insures the exact conservation of the
photon number and flux (in the explicit formulation) and automatically
switches from the optically thick to the optically thin regime. It scales
as $N\log N$ with the number of hydrodynamic resolution elements and is
independent of the number of sources of ionizing radiation (i.e.\
works equally fast for an arbitrary source function). 

We also describe an implementation of the algorithm in a Soften
Lagrangian Hydrodynamic code (SLH) and a multi--frequency approach
appropriate for hydrogen and helium continuum opacities. We present
extensive tests of our method for single and multiple sources in
homogeneous and inhomogeneous density distributions, as well as a
realistic simulation of cosmological reionization.
\end{abstract}

\end{frontmatter}

\def\kms{\mbox{\,km s$^{-1}$}}
\def\mpc{\mbox{\,Mpc}}
\def\Mpc{\mbox{\,Mpc}}
\def\kpc{\mbox{\,kpc}}
\def\pc{\mbox{\,pc}}
\def\Ms{\mbox{\,M$_{\odot}$}}
\def\Ls{\mbox{\,L$_{\odot}$}}
\def\cm{\mbox{\,cm}}
\def\ccc{\mbox{\,cm$^{-3}$}}
\def\erg{\mbox{\,erg}}
\def\s{\mbox{\,s}}
\def\yr{\mbox{\,yr}}
\def\Myr{\mbox{\,Myr}}

\def\tento#1{\times 10^{#1}}

\def\HH{\mbox{\,H$_2$}}

\def\K{{\rm \ K}}
\def\sr{{\rm \ sr}}
\def\eV{{\rm \ eV}}
\def\years{{\rm \ years}}
\def\yrs{{\rm \ years}}
\def\angstrom{\stackrel{o}{\rm A}}
\def\Hz{{\rm \ Hz}}
\def\nd#1{n_{ \rm #1}}
\def\k#1{k_{{\rm #1}}}

\def\fn#1{$^{\ref{#1}}$}
\def\fit#1{\footnotesize \it #1 }

\def\gtsima{$\; \buildrel > \over \sim \;$}
\def\ltsima{$\; \buildrel < \over \sim \;$}
\def\prosima{$\; \buildrel \propto \over \sim \;$}
\def\gsim{\lower.7ex\hbox{\gtsima}}
\def\lsim{\lower.7ex\hbox{\ltsima}}
\def\simgt{\lower.7ex\hbox{\gtsima}}
\def\simlt{\lower.7ex\hbox{\ltsima}}
\def\simpr{\lower.7ex\hbox{\prosima}}
\def\la{\lsim}
\def\ga{\gsim}
%
%
\def\cal{\it }
\def\A{{\cal A}}
\def\B{{\cal B}}
\def\ion#1#2{\rm #1\,\sc #2}
\def\HI{{\ion{H}{i}}}
\def\HII{{\ion{H}{ii}}}
\def\GI{{\ion{He}{i}}}
\def\GII{{\ion{He}{ii}}}
\def\GIII{{\ion{He}{iii}}}
\def\MH{{{\rm H}_2}}
\def\Hp{{{\rm H}_2^+}}
\def\Hm{{{\rm H}^-}}

\def\dim#1{\mbox{\,#1}}

\section{Introduction}

Numerous questions in physical cosmology and galaxy formation require
a detailed understanding of the physics of radiative transfer.  Of
particular interest are problems related to the reionization of the
intergalactic medium \citep[][and references herein]{GO97,HAR00,BL01}, 
absorption line
signatures of high redshift structures \citep{AM98,Kea99},
and hydrodynamic and thermal effects of UV emitting
central sources on galaxy formation \citep[e.g.][]{H95,SR98}.

Several attempts have been made to develop numerical algorithms for
solving the radiative transfer equation in astrophysics in general,
and in cosmology in particular. \citet{ANM99}
described a ray tracing approach for modeling radiative transfer
around point sources.  A related ray tracing approach was introduced
by \citet{RS99} which can also handle diffuse radiation
fields but finite computer memory limits the angular resolution and
therefore only poorly captures ionization fronts around point sources
on reasonably large grids. \citet{Cea01} discussed a Monte
Carlo approach to sample angles, energies and the time evolution of
the radiative transfer problem. A ray tracing scheme for use with
Smooth Particle Hydrodynamics was introduced by \citet{KB00}.  
All these approaches however suffer from the
excessive operation count and are costly to implement in realistic
simulations \citet[e.g.][]{SAH01}.

In order to implement a radiative transfer scheme into modern
numerical simulations, one needs to be able to solve the radiative
transfer equation in $O(N)$ operations (where $N$ is the number of
baryonic resolution elements, and we ignore a possible logarithmic
multiplier, so that $O[N\log N]$ still counts as O[$N$]). However, the
radiative transfer equation is six-dimensional in general and its
direct solution requires of the order of $O(N^{5/3}\times N_\nu)$
operations, where $N_\nu$ is the number of frequency bins (typically
must be about 200-300). Various formal solutions of the transfer
equation (like an attenuation equation) can be reduced to a four
dimensional (spatial plus frequency dimensions) form, but then the
calculation of the pairwise optical depth requires an operation count
of the order of $O(N\times N_S\times N_\nu)$, where $N_S$ is the
number of sources, which for an arbitrary source function can be equal
to $N$.  For some restricted case, like reionization by quasars, when
the number of sources $N_S$ is small enough, the whole solution can
indeed be computed directly in $O(N)$ operations \citep{ANM99}.
However, for more realistic cases the operation count increases
to $O(N^2)$.  In addition, in cosmological applications calculation of
the background radiation field would still require $N^{5/3}$
operations.

Thus, one has to use approximations to try to decrease the operation count.
The ``Local Optical Depth'' (LOD) approximation for the cosmological
radiative transfer was first introduced in \citet{GO97}
for the homogeneous distribution of sources, and later was
expanded on the case with an arbitrary source function in \citet{G00}.
This approximation is however highly approximate and uncontrolled.

Another possible approach is to use a so-called Variable Eddington Tensor
approximation by solving moment equations for the radiative field. Such
an approach allows to ensure exact conservation of photon numbers and
flux, and thus would automatically give correct speeds for the ionization
fronts. In this paper we describe such an approach whose main feature
is the Variable Eddington Tensor approximation with the tensor calculated
in the optically thin regime.

\citet{NPA98} first suggested to solve the moment
equations for the simulation of three dimensional radiative transfer
in cosmological hydrodynamics. They suggested to split the transfer of
radiation of point sources from the transfer of diffuse emitters. The
method presented here shares some of their ideas, but works with
different variables. We reformulate the equations of radiative
transfer by decomposing the radiation field into a uniform component
and a perturbation term. It is this new mathematical framework that we
will present in the following section. In Section~3 we discuss our
treatment of multiple frequency bins and also introduce a new
``Reduced Speed of Light'' approximation which allows us to overcome
numerous numerical problems. We then go on in Section~4 to introduce
particular details of our implementation of the algorithm. Section~5
is devoted to various one and three dimensional test cases that
exemplify the strengths and weaknesses of our approach. We also
present in that section a realistic cosmological simulation based on
our new scheme, and compare our scheme with the LOD approach.  We
conclude in section~6 and give a discussion of various performance
issues.

\section{Formalism}

\subsection{Basic equations}

The evolution of the specific intensity
$J_\nu$
in the 
expanding universe is given by the following equation:
\begin{equation}
        {\partial J_\nu\over\partial t} +
        {\partial\over\partial x^i}\left(\dot{x}^i J_\nu\right) -
        H\left(\nu{\partial J_\nu\over\partial\nu}-3J_\nu\right) =
        - k_\nu J_\nu + S_\nu.
        \label{Jnueq}
\end{equation}  
Here $x^i$ are the comoving coordinates, $H$ is the Hubble constant,
$k_\nu$ is the absorption coefficient, $S_\nu$ is the source function,
and $\dot{x}^i=cn^i/a$, where $n^i$ is the unit vector in the direction of
photon propagation.

Equation (\ref{Jnueq}) includes all relativistic effects and therefore is quite
complex. But in the same spirit as equations for particle motions in the
expanding universe are split into a uniform relativistic background and
Newtonian perturbations, we can split equation (\ref{Jnueq}) into a 
background and perturbations \citep{GO97}.

First, we define the mean specific intensity,
\begin{equation}
        \bar J_\nu(t) \equiv \langle J_\nu(t,\vec{x},\vec{n})\rangle,
        \label{Jnubardef}
\end{equation}
where the averaging operator $\langle\rangle$ acting on a function
$f(\vec{x},\vec{n})$ of position and direction is defined as:
\begin{equation}
        \langle f(\vec{x},\vec{n})\rangle =
        \lim_{V\rightarrow\infty} {1\over 4\pi V}
        \int_V d^3x \int d\Omega f(\vec{x},\vec{n}).
        \label{averagoper}
\end{equation}
The mean specific intensity (``background'') satisfied the following
equation:
\begin{equation}
        {\partial \bar J_\nu\over\partial t} -
        H\left(\nu{\partial \bar J_\nu\over\partial\nu}-3\bar J_\nu\right) =
        - \bar k_\nu \bar J_\nu + \bar S_\nu,
        \label{Jnubareq}
\end{equation}
where, {\it by definition\/}, 
\[
        \bar S_\nu \equiv \langle S_\nu \rangle,
\]
and
\[
        \bar k_\nu \equiv {\langle k_\nu J_\nu\rangle\over \bar J_\nu}.
\]
It is important to emphasize here that we use 
the overbar symbol
merely as a part
of notation and not to denote the space average; in particular, 
$\bar k_\nu$ is not a space average of $k_\nu$ since it is 
weighted by the local value of the specific intensity $J_\nu$.

We can now introduce a new quantity, which we call relative (to the mean) 
specific intensity $f_\nu(t,\vec{x},\vec{n})$,
which quantifies the
difference between the local value of the specific intensity
and the mean,
\begin{equation}
        J_\nu \equiv f_\nu\bar J_\nu,
        \label{fnudef}
\end{equation} 
so that $\langle f_\nu\rangle=1$. In the dynamic analogy, $f_\nu$ corresponds
to $1+\delta$, where $\delta$ is the cosmic overdensity.

It is now straightforward
to derive the following equation for $f_\nu$:
\begin{equation} 
        {\partial f_\nu\over\partial t} + 
        {\partial\over\partial x^i}\left(\dot{x}^i f_\nu\right) =
        H\nu{\partial f_\nu\over\partial\nu} - 
        (k_\nu-\bar k_\nu+{\bar S_\nu\over \bar J_\nu}) f_\nu + 
	{S_\nu\over \bar J_\nu}.
        \label{fnueqfull}
\end{equation}
This equation is of course not simpler than the original equation
(\ref{Jnueq}). But, if 
we restrict ourselves to
scales significantly smaller than the horizon size, and matter velocities
much smaller than the speed of light (Newtonian limit), then we can
simplify equation (\ref{fnueqfull}) substantially by noting that the
relative specific intensity does not change substantially and the universe
does not expand substantially over the period of time a photon needs to travel
over the scale under consideration (for example, a size of a computational 
box). Therefore, we can ignore the derivative 
with respect to the frequency. Strictly speaking, we should also discard
the time derivative, as it has the same power of $v/c$ as the frequency
derivative. However, we will retain it for a moment, keeping in mind that
it is small - as we explain below, the time derivative turns out to be
quite useful for the numerical implementation of our scheme.

Under these assumptions, equation (\ref{fnueqfull}) reduces to the following
simple equation:
\begin{equation}
        {a\over c}{\partial f_\nu\over\partial t} +
	n^i{\partial f_\nu\over\partial x^i} = 
        - \hat\kappa_\nu f_\nu + \psi_\nu,
        \label{fnueq}
\end{equation}
where
\[
	\kappa_\nu \equiv {a\over c}k_\nu,
\]
\begin{equation}
	\hat\kappa_\nu \equiv \kappa_\nu - \bar\kappa_\nu +
	{\bar S_\nu\over \bar J_\nu}
	\label{kappadef}
\end{equation}
and
\[
	\psi_\nu \equiv {a\over c}{S_\nu\over \bar J_\nu}.
\]

Equations (\ref{Jnubareq}) and (\ref{fnueq}) thus separate the average 
cosmological evolution of specific intensity (including redshift
and effects of ``distant'' sources - sources outside the simulation box)
from the local evolution due to inhomogeneous absorption and sources inside
the box.

\subsection{Moments of the transfer equation}

The transfer 
equation (\ref{fnueq}) is a six-dimensional equation in the phase space,
and as such offers a considerable difficulty to solve numerically.
Pursuing the analogy with dynamics further, we can define moments of
the radiation field similarly to how density, momentum, pressure etc of
a medium 
are defined as moments of the distribution function. Specifically,
we define the (relative to the mean) radiation energy density $E_\nu$,
\[
	E_\nu(t,\vec{x}) \equiv {1\over 4\pi}\int d\Omega\, f_\nu(t,\vec{x},\vec{n}),
\]
the flux $F^i_\nu$,
\[
	F^i_\nu(t,\vec{x}) \equiv {1\over 4\pi}\int d\Omega\, n^i f_\nu(t,\vec{x},\vec{n}),
\]
and the so-called Eddington tensor $h^{ij}_\nu$,
\[
	E_\nu(t,\vec{x}) h^{ij}_\nu(t,\vec{x}) \equiv 
	{1\over 4\pi}\int d\Omega\, n^i \vec{n} f_\nu(t,\vec{x},\vec{n}).
\]
In particular, 
\[
	{\rm Tr\,}h^{ij}_\nu = 1\ \ \ \mbox{and}\ \ \
	\langle E_\nu \rangle = 1.
\]

Taking moments of equation (\ref{fnueq}), we can derive equations for the
moments of the radiation field (assuming that the source function
is isotropic):
\begin{equation}
	{a\over c}{\partial E_\nu\over\partial t} +	
	{\partial F^i_\nu\over \partial x^i} = - \hat\kappa_\nu E_\nu
	+ \psi_\nu
	\label{eeq}
\end{equation}
and
\begin{equation}
	{a\over c}{\partial F^j_\nu\over\partial t} +	
	{\partial \over \partial x^i} E_\nu h^{ij}_\nu= - 
	\hat\kappa_\nu F^j_\nu.
	\label{feq}
\end{equation}

It is easy to see that equations (\ref{eeq}) and (\ref{feq}) are
in a conservative form, and so represent the conservation of the number of
photons and the flux respectively.

Equations (\ref{eeq}) and (\ref{feq}) constitute a system of two partial
differential equations. The system is however not closed since the Eddington
tensor is not specified by these two equations. However, {\it if the Eddington 
tensor is given as a function of position and time\/}, the system becomes
closed. In the following section
we assume that the Eddington tensor is known and describe the numerical
implementation of an algorithm based on equations (\ref{eeq}) and 
(\ref{feq}). We then discuss how to define the Eddington tensor, thus
completing the numerical scheme.

\section{Specific implementation}

\subsection{The ``Reduced Speed of Light'' approximation}
\label{rslsec}

In this section we omit the frequency dependence for clarity. Equations
(\ref{eeq}) and (\ref{feq}) can be combined into one equation preserving
the accuracy of the Newtonian approximation. Namely, 
from (\ref{feq}) we have:
\[
	F^j = -{1\over \hat\kappa} \left({a\over c}{\partial F^j\over
	\partial t} +	
	{\partial E h^{ij}\over \partial x^i} \right).
\]
Substituting this into (\ref{eeq}), we obtain:
\begin{equation}
	{a\over c}{\partial E\over\partial t} +	
	 \hat\kappa E - \psi -
	{\partial \over \partial x^j}\left( {1\over \hat\kappa} 
	{\partial E h^{ij}\over \partial x^i} \right) = 
	{a\over c} {\partial \over \partial x^j}
	\left({1\over \hat\kappa }{\partial F^j\over
	\partial t}\right).
	\label{eqaa}
\end{equation}
The right hand side can be expanded into two terms,
\[
	{a\over c} {\partial \over \partial x^j}
	\left({1\over \hat\kappa }{\partial F^j\over
	\partial t}\right) =
	-{a \hat\kappa_{,j}\over c\hat\kappa^2} {\partial F^j\over
	\partial t} +
	{a\over c\hat\kappa} {\partial \over \partial t}
	{\partial F^j\over
	\partial x^j}.
\]
Both of these terms are of the order of $1/c$, and thus can be dropped in
the Newtonian limit. We therefore neglect the first term, but retain the second
one, keeping in mind that it is of the higher order in $1/c$ than the
leading terms in equation (\ref{eqaa}). That leaves the right hand side of
equation (\ref{eqaa}) in the following form:
\[
	{a\over c\hat\kappa} {\partial \over \partial t}
	\left(	-{a\over c}{\partial E\over\partial t}
	- \hat\kappa E
	+ \psi \right),
\]
of which we retain only one term,
\[
	-{a\over c} {\partial E\over \partial t},
\]
with the same $1/c$ accuracy. This reduces the original equation
(\ref{eqaa}) to the following final form:
\begin{equation}
	2{a\over c}{\partial E\over\partial t} = 	
	{\partial \over \partial x^j}\left( {1\over \hat\kappa} 
	{\partial E h^{ij}\over \partial x^i} \right) 
	- \hat\kappa E + \psi.
	\label{difeq}
\end{equation}
This one equation is equivalent to the original two equations
(\ref{eeq}) and ({feq}) in the Newtonian limit, and the term with
the time derivative is small and formally of the order of $1/c$.

The physical reason for retaining the time derivative and throwing away
other terms which are formally of the same order is that at the ionization
front $E$ can change very rapidly making this term significantly larger
than the terms we neglected, even if formally they have the same $1/c$ order.

In the Newtonian limit the speed of light is infinite, which in practice
means that we ignore all the terms of the order of $v/c$. Thus, it actually
does not matter what the specific value of the speed of light is as long as
$v/c\ll1$, and we can replace the true speed of light $c$ in equation
(\ref{difeq}) with some other quantity $\hat{c}$ without violating the
validity of the Newtonian limit as long as $v/\hat{c}\ll1$. For example,
for modeling cosmological reionization we can adopt a value for $\hat{c}$
as low as $1,000\dim{km}/\dim{s}$, because typical gas velocities during 
reionization do not exceed $100\dim{km}/\dim{s}$.

This trick of ``reducing the speed of light'' (which we call the Reduced
Speed of Light, or RSL, approximation) allows us to achieve a very
important goal. Namely, if we ignore the time derivative term in 
equation (\ref{difeq}), the equation turns elliptic, which is difficult
to solve for an arbitrary $\hat\kappa$. Thus, keeping the
time derivative in equation (\ref{difeq}) allows us to solve this equation
as a dynamical equation and use simple stable explicit schemes.

There is however one very serious limitation - the time-step in any explicit
scheme will be limited by an appropriate Courant condition to a value which
is about $c_S/c$ times smaller than the hydrodynamic time-step (here $c_S$
is the gas sound speed). For the intergalactic gas at $10,000\dim{K}$ this
ratio can be as small as $1/30,000$. Clearly, this will make such a scheme
impractical, as it is impossible to make 30,000 radiative transfer time steps
for every hydrodynamic time step in a realistic simulation.

The RSL approximation allows to reduce this ratio dramatically to less than
100, and in our tests described below we even find that taking 
$\hat{c}\sim c_S$ does not lead to large errors.

Thus, in the RSL approximation equation (\ref{eqaa}) reduces to the following
form:
\begin{equation}
	2{a\over \hat{c}}{\partial E\over\partial t} = 	
	{\partial \over \partial x^j}\left( {1\over \hat\kappa} 
	{\partial E h^{ij}\over \partial x^i} \right) 
	- \hat\kappa E + \psi
	\label{rsl}
\end{equation}
where $\hat{c}\leq c$.

It is important to note here that the reduction in the speed of light 
{\it does not\/} limit the propagation speed of an ionization front
to $\hat{c}$. Equation (\ref{rsl}) is a diffusion equation, which
has an infinite speed of signal propagation. In other words, the
RSL approximation  puts no limit on the speed of ionization 
fronts and thus 
can sometimes lead to unphysical, faster than the speed of light
propagation of the ionization fronts \citep{ANM99}. In practice, we
put a limit on the maximum possible speed of an ionization front by
limiting the number of time-steps. This is discussed in greater
detail in \S\ref{ifssec}.

\subsection{Multi-frequency transfer}

Equation (\ref{rsl}) is a diffusion equation, and as such is not more
difficult to solve than equations of hydrodynamics (in fact, it is much
easier to solve). However, we have omitted the frequency dependence in
the derivations of the previous section. Tests indicate that in order
to calculate the various ionization and heating rates with sufficient
precision on a logarithmically spaced mesh in frequency space, at least
20 points per e-folding are required, which translates into at least 200
frequency bins. Solving equation (\ref{rsl}) 200 times at each hydrodynamic
time step would be too
prohibitive with the currently available computer resources.

We therefore need to introduce further approximation in order to break this
unfavorable scaling. In order to do that, it is instructive to consider the
optically thin case first.
In this regime the angle averaged solution to
equation (\ref{Jnueq}) is (omitting time dependence for clarity):
\[
	\langle J_\nu\rangle_\Omega
	(\vec{x}) = \bar{J}_\nu + {a\over 4\pi c} \int d^3 x_1 
	{S_\nu(x_1^i)-\bar{S}_\nu\over (\vec{x}-\vec{x}_1)^2}.
\]
Both, $\bar{J}_\nu$ and $S_\nu$ may have very complicated frequency
dependence, which would require a large number of frequency bins.
However, if we
assume that all sources have the same frequency
dependence, i.e.
\[
	S_\nu(t,\vec{x}) = g_\nu \rho_*(t,\vec{x}),
\]
where $g_\nu$ depends only on frequency, and $\rho_*$ depends only on time and
space (say, the mass density of stars), then in the optically thin regime
$J_\nu$ as a function of frequency separates into two terms:
\begin{equation}
	\langle J_\nu\rangle_\Omega = \bar{J}_\nu E_1 + g_\nu D_2
	\label{fnsor}
\end{equation}
where $E_1 = 1$ and 
\begin{equation}
	D_2 = {a\over 4\pi c} \int d^3 x_1 {\rho_*(\vec{x}_1)-\bar\rho_*\over 
	(\vec{x}-\vec{x}_1)^2},
	\label{dtot}
\end{equation}
and both $E_1$ and $D_2$ are independent of frequency. Thus, in this regime
the frequency dependence is factorized in two one dimensional functions
$\bar{J}_\nu$ and $g_\nu$, and all relevant ionization and heating rates
need to be computed only once and then simply rescaled by frequency-independent
factors $E_1$ and $D_2$.

This procedure is equivalent to separating all sources into ``distant''
ones, which include full redshift dependence and cosmological terms,
as well as diffuse radiation, and whose frequency dependence is described
by $\bar{J}_\nu$,  and into ``near'' sources which are
just Newtonian perturbations on the GR background with the frequency
dependence given by $g_\nu$.

In a general case, let's try to use ansatz (\ref{fnsor}). In that case
$E_1$ and $D_2$ are no longer frequency-independent, but we will
consider them being ``weakly frequency dependent'' in some sense. 
We then proceed by 
approximating their frequency dependence with ``effective column
densities'',
\[
	Q_\nu = Q_{\rm OT} \exp\left(-\sum_\alpha \sigma_\nu^{(\alpha)}
	N_{\rm eff}^{(\alpha)} \right),
\]
where $Q$ is either $E_1$ or $D_2$, and $Q_{\rm OT}$ is the respective
quantity in the optically thin regime. Index $\alpha$ runs over the
list of species, which normally includes \HI, \GI, and \GII, but
may also include other species, for example such as $\MH$ or dust.
\footnote{In practice we do a slightly more
complicated trick, and assume that $N_{\rm eff} = N_1 + X N_2$ where
$X$ is the abundance of a species $\alpha$ (which can change during
the hydrodynamic time-step), and $N_{1,2}$ are now fixed
functions of position during the time-step (independent of time during the
hydrodynamic time-step). The quantity $N_2$ is defined as
$N_2=n L$, where $n$ is the number density of a given species at a given point,
and $L$ is the size of the resolution element. The latter quantity is not
defined precisely, but rather up to a factor of a few, but the numerical scheme
is quite insensitive to its specific value: we varied it by a factor of 10 
with little change to the final solution.}
Any rate of a specific physical process (such as photoionization rates,
photoheating rates, etc) will now depend on those $N_{\rm eff}^{(\alpha)}$.
For the case of just three species (say, $\alpha$=\HI,\GI,\GII), we
precompute these rates
with fine frequency dependence of $\bar{J}_\nu$ and $g_\nu$, store
them in a 3D table, and then simply pick up a number from a table when it is
needed. For four or more species rates need to be recalculated ``on the
fly'' because the four-dimensional table is not practical.
Tests show that precomputing gives about a factor of 10-30 speed up.

In order to use ansatz (\ref{fnsor}), we need to define the two quantities
$E_{1,\nu}$ and $D_{2,\nu}$.  First, we note that since 
\[
	\langle J_\nu\rangle_\Omega = \bar{J}_\nu E_\nu,
\]
then
\[
	E_\nu = E_{1,\nu} + {g_\nu\over\bar{J}_\nu} D_{2,\nu}.
\]

The simplest way to split the ``full'' function $E$ into $E_1$ and
$D_2$ is by the following two equations (we put tildes over $E_1$ and $D_2$
because, albeit the simplest, this scheme does not work in practice):
\begin{eqnarray}
	2{a\over \hat{c}}{\partial \tilde E_{1,\nu}\over\partial t} & = &
	{\partial \over \partial x^j}\left( {1\over \hat\kappa_\nu} 
	{\partial \tilde E_{1,\nu} h_{1,\nu}^{ij}\over \partial x^i} \right) 
	- \hat\kappa_\nu \tilde E_{1,\nu} + \bar\psi, \nonumber \\
	2{a\over \hat{c}}{\partial \tilde D_{2,\nu}\over\partial t} & = &
	{\partial \over \partial x^j}\left( {1\over \kappa_\nu} 
	{\partial \tilde D_{2,\nu} h_{2,\nu}^{ij}\over \partial x^i} \right) 
	- \kappa_\nu \tilde D_{2,\nu} + \rho_* - \bar\rho_*,
	\label{schone}
\end{eqnarray}
where $h_{1,\nu}^{ij} = \delta^{ij}/3$ and $h_{2,\nu}^{ij}$ 
is the original Eddington tensor ($h_\nu^{ij}$ from eq.\ [\ref{feq}]).
Note, that the second equation now contains $\kappa_\nu$ and not 
$\hat\kappa_\nu$ because the $\bar{J}_\nu$ dependence was factored
out.

However, the scheme (\ref{schone}) does not work, because $\tilde E_{2,\nu}=
g_\nu\tilde{D}_2/\bar{J}_\nu$ can then be
negative and positive, and numerical truncation errors will prevent
the exact cancellation for $E_\nu=\tilde E_{1,\nu}+\tilde E_{2,\nu}$
outside the ionization front, where $E_\nu$ should be exactly zero.
Therefore, we use a different splitting:
\begin{eqnarray}
	2{a\over \hat{c}}{\partial E_{1,\nu}\over\partial t} & = &
	{\partial \over \partial x^j}\left( {1\over \hat\kappa_\nu} 
	{\partial E_{1,\nu} h_{1,\nu}^{ij}\over \partial x^i} \right) 
	- \hat\kappa_\nu E_{1,\nu} + \bar\psi\Lambda_\nu,\nonumber \\
	2{a\over \hat{c}}{\partial D_{2,\nu}\over\partial t} & = & 	
	{\partial \over \partial x^j}\left( {1\over \kappa_\nu} 
	{\partial D_{2,\nu} h_{2,nu}^{ij}\over \partial x^i} \right) 
	- \kappa_\nu D_{2,\nu} + \rho_* - \bar\rho_*\Lambda_\nu.
	\label{rslab}
\end{eqnarray}
Quantity $\Lambda_\nu$ is not specified here, it is only required to
go to 0 when the mean free path is much shorter than the size of the
periodic computational box, and to 1 when the means free path is
much larger than the box size. Irrespective of the choice of
$\Lambda_\nu$, periodic boundary conditions of the computational box
will guarantee that our solution will deviate from the solution in the
exact case of non-periodic universe when the mean free path is close
to the size of the computational box. Therefore, the specific choice
for $\Lambda_\nu$ is not that important. In this paper we adopt
$\Lambda_\nu = \exp(-\langle\kappa_\nu\rangle{\cal L})$, where
\[
	{\cal L} = {1\over 4\pi} \int { d^3 x_1 \over 
	(\vec{x}-\vec{x}_1)^2}
\]
(the latter integral is apparently independent of $\vec{x}$). We verified
that other choices of $\Lambda_\nu$ give different solutions when the mean
free path is close to the box size, but converge to the same solution
when the mean free path is much shorter or much larger than the box size.

We emphasize that this procedure is only required in order to reduce
the number of frequency bins in the calculation. It can be entirely
circumvented if the full frequency space coverage can be afforded.

The scheme presented above can be easily generalized on a case when there
are several kinds of sources, each with its own spectral shape $g_\nu$.

\subsection{Numerical implementation}

Equations (\ref{rslab}) are then solved by the following 
simple semi-explicit scheme. Let $Q$ be either $E_1$ or $D_2$
(we omit the frequency dependence in this section for the sake of clarity),
\[
	\Delta_2 \equiv {\partial \over \partial x^j}\left( {1\over \kappa} 
	{\partial  Q h^{ij}\over \partial x^i} \right) 
\]
with the appropriate tensor $h^{ij}$, $P$ is the source term
($\bar\psi\Lambda$ for $E_1$ or $\rho_* - \bar\rho_*\Lambda_\nu$ for $D_2$), 
and 
$d\xi=\hat{c}dt/(2a)$. Then equations
(\ref{rslab}) are shorthanded as
\begin{equation}
	{\partial Q\over\partial\xi} = 
	\Delta_2 - \kappa Q + P.
	\label{eqa}
\end{equation}
Let the superscript $(n)$ label the $n$-th time step in the time-discretized
version of equation (\ref{eqa}). We then adopt a semi-explicit scheme for
solving equation (\ref{eqa}), in which we treat the Laplacian and the source
term explicitly, and absorption implicitly (since the time scale for absorption
can be extremely short),
\[
	{\tilde{Q}^{(n+1)} - Q^{(n)}\over \delta\xi} = 
	\Delta_2^{(n)} - \kappa^{(n)} Q^{(n+1)} + P^{(n)},
\]
which can be reduced to the explicit expression for $\tilde{Q}^{(n+1)}$,
\begin{equation}
	\tilde{Q}^{(n+1)} = Q^{(n)} + \delta Q^{(n)},
	\label{dirsch}
\end{equation}
and
\[
	\delta Q^{(n)} \equiv \delta\xi{\Delta_2^{(n)} - 
	\kappa^{(n)} Q^{(n)} + P^{(n)}\over1+\delta\xi\kappa^{(n)}}.
\]
The time step $\delta\xi$ is determined from the standard Courant condition
for the diffusion equation,
\[
	\delta\xi = C \min({3\over2}\Delta x^2 \kappa),
\]
where $\Delta x$ is the mesh cell size, and $C=0.9$ is the Courant number.

Our scheme is implemented in the Soften Lagrangian Hydrodynamics (SLH) code
\citep{G95,GB96,G00}. 
The SLH code follows all physical quantities on a
moving deformed mesh specified by a coordinate transformation from the
quasi-Lagrangian space $q^k$ (where the mesh is uniform) into the real
space $x^i=x^i(t,q^k)$. In the quasi-Lagrangian space the Laplacian can
be represented by using the deformation tensor
\[
        A^i_k \equiv {\partial x^i \over \partial q^k},
\]
its inverse $B^k_i$, and its determinant 
$\A$:
\[
	\Delta_2[Q] = 
	{1\over \A} {\partial \over \partial q^k} \left[
        {1\over\kappa+\Delta\kappa} \A B^k_i B^l_j 
        {\partial Q h^{ij}\over \partial q^l}
        \right],
\]
where we also introduced ``artificial viscosity'' 
$\Delta\kappa\equiv\epsilon/\Delta x$ to prevent the denominator from
becoming zero. The results are insensitive to the choice of $\epsilon$ as long
as it is less than about 0.03.
The cell size $\Delta x$ is now defined as
\[
	\Delta x^2=3\left(B_i^kB_j^l\delta_{ij}\delta^{kl}\right)^{-1}.
\]
	
The description above completes our numerical scheme except for the choice
of the ``reduced speed of light'' $\hat{c}$. We determine it from the
following condition:
\begin{equation}
	\hat{c} = c\times \min(1,N_{\rm max}\delta t/\Delta t)
	\label{chat}
\end{equation}
where $\Delta t$ is the hydrodynamic time-step during which equations
(\ref{rslab}) are solved, $\delta t$ is the time-step for equations
(\ref{rslab}) determined from the Courant condition, and $N_{\rm max}$
is the upper limit on the number of time-steps in one hydro time-step.
This is equivalent to solving equations (\ref{rslab}) in $N_{\rm max}$
time steps at most - if more time steps is required, the integration
is interrupted at the moment when the maximum allowed 
number of time steps is reached. Our scheme is fairly insensitive to
the choice of $N_{\rm max}$, as discussed in the next subsection.

Thus, the whole scheme for solving RT within the SLH code is the following. 
At each hydrodynamic 
time step:
\begin{itemize}
\item gravity and hydro are updated.
\item equations (\ref{rslab}) are solved for $N_{\rm max}$ steps at four
frequency bins ($\nu$=0,\HI,\GI,\GII). Then $N^{(\alpha)}_{\rm eff}$
are calculated.
\item equations for gas temperature and ionization state are solved given 
	$N^{(\alpha)}_{\rm eff}$.
\end{itemize}

\subsection{Signal propagation speed}
\label{ifssec}

As we discussed in \S\ref{rslsec}, the RSL approximation by itself
does not limit the signal propagation speed. However, an adopted limit
on the number of time-steps $N_{\rm max}$ does limit the speed
with which a signal can propagate across the mesh. Namely, since we
are using a 3-point finite differencing scheme for computing the
Laplacian, in one time-step a signal can only propagate one cell
on our quasi-Lagrangian mesh. Let us first assume that our mesh is
uniform in real space as well, i.e.\ that our quasi-Lagrangian
space coincides with the real space. In that case during one hydrodynamic
time-step $\Delta t$ the signal can only travel the physical distance
$\Delta r = a L N_{\rm max}/N_{\rm box}$, where $L$ is the comoving 
size of the computational box, and $N_{\rm box}$ is the number of cells
along one dimension of the box. Thus, the propagation speed of a signal
over the box is limited to
\[
	v_{\rm max} = {a L N_{\rm max}\over N_{\rm box}\Delta t},
\]
or
\begin{equation}
	{v_{\rm max}\over c} = {a L_3 N_{\rm max}\over N_{\rm box}
	\Delta t_{10}},
	\label{vmax}
\end{equation}
where $L_3\equiv L/3\dim{Mpc}$ and $\Delta t_{10}=\Delta t/10\dim{Myr}$.
For example, for the test shown in Fig.\ \ref{figHS} this number is about
$1.2N_{\rm max}$. The test was performed with $N_{\rm max}=1$ and
$N_{\rm max}=4$ and we found no noticeable difference between the two
cases, which is not surprising since the top speed of the ionization
front in this case is only about $4000\dim{km/s}$. This speed is however
much larger than $N_{\rm max}$ times the sound speed in the gas
($16\dim{km/s}$ behind the ionization front and $4\dim{km/s}$ outside
it), which illustrates that the RSL approximation does not limit
the propagation speed of the signal.

Thus, one can in principle adjust $N_{\rm max}$ in equation (\ref{vmax})
so that the signal propagation speed is always fixed to $c$ (unless
it would require $N_{\rm max}<1$). However, for realistic cosmological 
simulations presented in \S\ref{sim} we do not do that, because
the SLH code allows for severe deformations of the computational
mesh in real space, and in this case
the signal propagation speed is only fixed in the quasi-Lagrangian
space for a fixed value of $N_{\rm max}$, but varies at different 
spatial locations in real space. Instead, we keep
$N_{\rm max}$ fixed throughout the simulation, and we have tried
two values, $N_{\rm max}=20$ and $N_{\rm max}=100$, which give
visually indistinguishable results. 
Incidentally, during the whole simulation with $N_{\rm max}=20$
the signal propagation speed $v_{\rm max}$ never deviated from $c$
by more than a factor of 10 (and it is respectively 5 times larger
for $N_{\rm max}=100$).

\subsection{The choice of the Eddington tensor}

The scheme presented in the previous sections is complete given the
Eddington tensor $h_\nu^{ij}$. The devil of course is in the choice
for the Eddington tensor, because it allows to close the moment
hierarchy. 

The Eddington tensor can be written in a closed form as
\begin{equation}
	h_\nu^{ij} = {P_\nu^{ij}\over {\rm Tr\,}P_\nu^{ij}},
	\label{ouret}
\end{equation}
where
\begin{equation}
	P_\nu^{ij} = \int d^3 x_1 \rho_*(\vec{x}_1) 
	e^{-\tau_\nu(\vec{x},\vec{x}_1)}
	{(x^i-x_1^i)(x^j-x_1^j)\over (\vec{x}-\vec{x}_1)^4},
	\label{expij}
\end{equation}
where $\tau_\nu(\vec{x},\vec{x}_1)$ is the optical depth between
points $\vec{x}$ and $\vec{x}_1$. Evaluation of this integral requires
$O(N^{5/3})$ operations, and thus this form is unsatisfactory. Thus,
we need to make a further approximation to achieve a favorable
$O(N)$ scaling.
However, this approximation has to be introduced carefully so that
the physical meaning of equation (\ref{expij}) is not violated. Specifically,
equation (\ref{expij}) describes the fact that radiation propagates
from a set of discrete sources along straight lines, and any new
approximation has to preserve this property. 

In this paper we propose to use Eddington tensors calculated
in the optically thin regime,
\begin{equation}
	P^{ij} = \int d^3 x_1 \rho_*(\vec{x}_1)
	{(x^i-x_1^i)(x^j-x_1^j)\over (\vec{x}-\vec{x}_1)^4}
	\label{otpij}
\end{equation}
(in this case $P^{ij}$ and $h^{ij}$ become frequency independent).
The latter integral can now be computed in $O(N)$ operations by
standard algorithms used to compute high-resolution gravitational
forces such as P$^3$M, Tree, or Adaptive Mesh Refinement.

It is important to underscore that in our approximation the
physical meaning of equation (\ref{expij}) is preserved - radiation
still propagates from a set of discrete sources along straight lines -
and only the relative weight of different sources is misestimated. The
latter however does not mean that the number of ionizing photons
emitted by a given source is misestimated as well. By separating the
computation of the Eddington tensor from the first two moments of the
transfer equation, we in effect separate two physical processes:
absorption and emission of photons (which is described by the two
moments) and propagation of photons along the straight lines from the
source locations (which is described by the Eddington tensor). The
first we compute almost exactly (subject to the RSL approximation and
finite number of frequency bins), while the second we compute in the
optically thin regime. In other words, we insure the exact
conservation of the number and the flux of photons, but may
occasionally advect some flux in the wrong direction.

We also stress that out approximation differs significantly from other
schemes, like the diffusion approximation.  In the latter the
Eddington tensor is computed locally, and it thus misses the
non-locality present in equations (\ref{expij}) and (\ref{otpij}). For
example, the diffusion approximation will grossly fail for the case
when a small \HII\ region is overrun by a larger \HII\ region from a
more distant but a much stronger source, whereas our approach will
give a quite accurate representation of the radiation field in that
case (this test is discussed in the next section).

\section{Testing the scheme}

\subsection{Single source tests}

In order to verify the accuracy of our method, we have compared the results
obtained with the 3D code to the predictions of the exact solution in
spherical symmetry.

In spherical symmetry equation (\ref{fnueq}) with no time derivative term and
for a point source at the
center ($\psi_\nu=\Psi_\nu\delta[\vec{x}]$) has an exact solution,
\[
	f_\nu(t,r,\vec{n}) = 
	{\Psi_\nu\over 4\pi r^2} 
	e^{\displaystyle-\int_0^r \hat\kappa_\nu(t,r^\prime)
	dr^\prime} \delta(\vec{n}-\vec{e}_r).
\]
Our implementation of a spherically symmetric code has 1000 points per decade in
the radial direction (which has been verified to be sufficient to converge to
1\% accuracy) and exactly the same frequency dependence and ionization
and heating rates as the 3D SLH code.

\begin{figure}
\includegraphics[width=\hsize]{\figdir/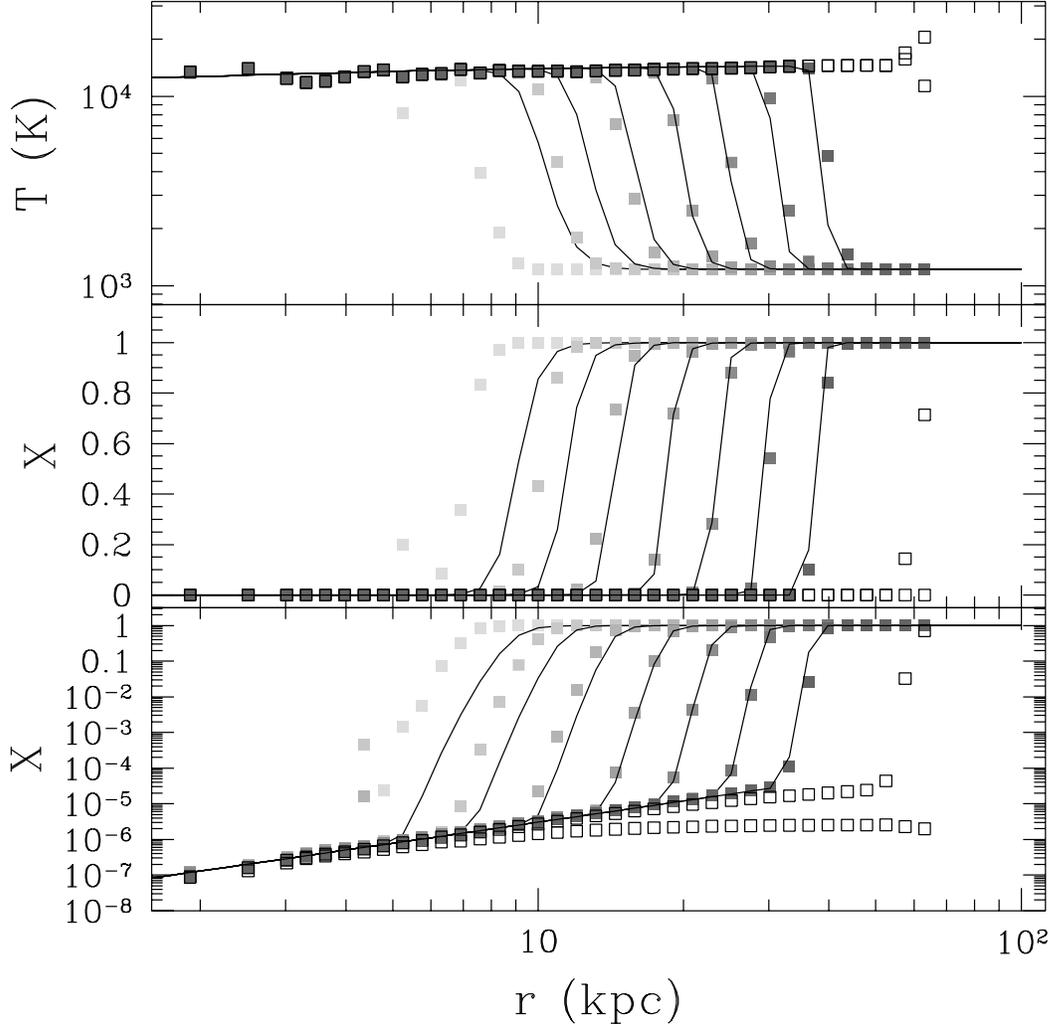}
\caption{\label{figHS}Propagation of the ionization front into the uniform 
neutral medium with $n=2.5\times10^{-5}\dim{cm}^{-3}$ from a point source
with $\dot{N}_{\rm ph}=10^{52}\dim{s}^{-1}$. Solid lines show the exact
spherically symmetric solution, whereas grey symbols show the 3D approximate
solution after 3, 6, 12, 25, 50, 100, and 200 hydrodynamic time steps
(in the order of the increasing darkness of a symbol). Two sets of open symbols
show the 3D solution after 700 and 1000 hydrodynamic time steps. The latter
moment corresponds to the time when the entire computational box is optically
thin. The upper panel shows the gas temperature, the middle panels shows
the hydrogen ionization fraction on a linear scale, and the bottom panel
shows the hydrogen ionization fraction on a logarithmic scale.}
\end{figure}
Figure \ref{figHS} now illustrates how our scheme works for a case of a
point source in a uniform medium. In this test the point source is located
at the center of a $64^3$ mesh in a $70\dim{kpc}$ box with periodic boundary
conditions.
The source spectrum contains only hydrogen 
ionizing photons. We notice that the approximate 3D solution converges to
the exact solution in about 10 hydrodynamic time steps, after which time
the 3D scheme reproduces accurately not only the position of the ionization 
front, but also its detailed temperature and ionized fraction profiles.
The deviation at the initial moment is most likely due to finite resolution
of the 3D code.
At later times the ionization front leaves the computational box and the
solution properly approaches the optically thin regime (open symbols in
Fig.\ \ref{figHS}) when most of the box has a uniform radiation background
and a proximity region exists around the source.

\begin{figure}
\includegraphics[width=0.45\hsize]{\figdir/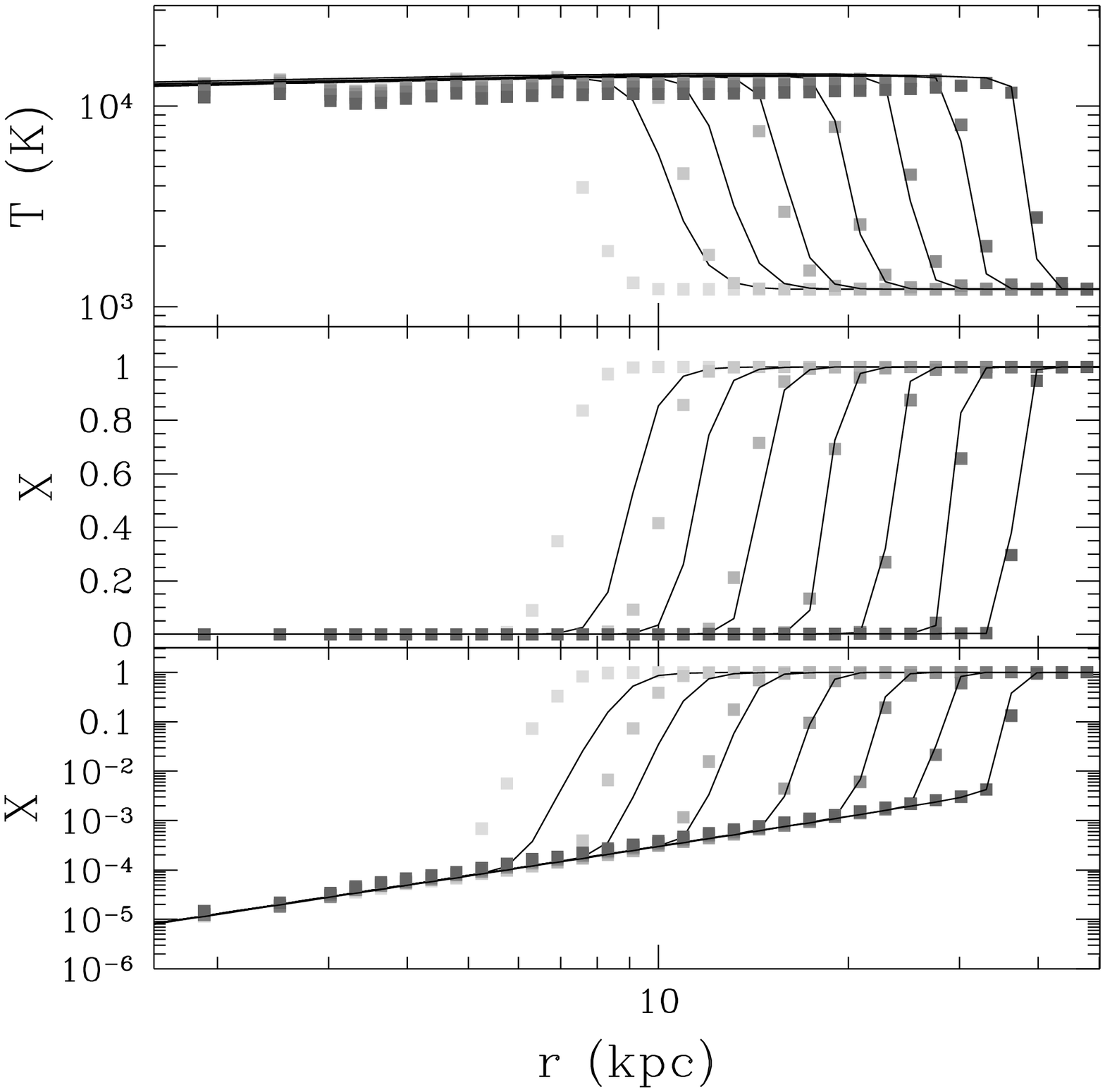}
\includegraphics[width=0.45\hsize]{\figdir/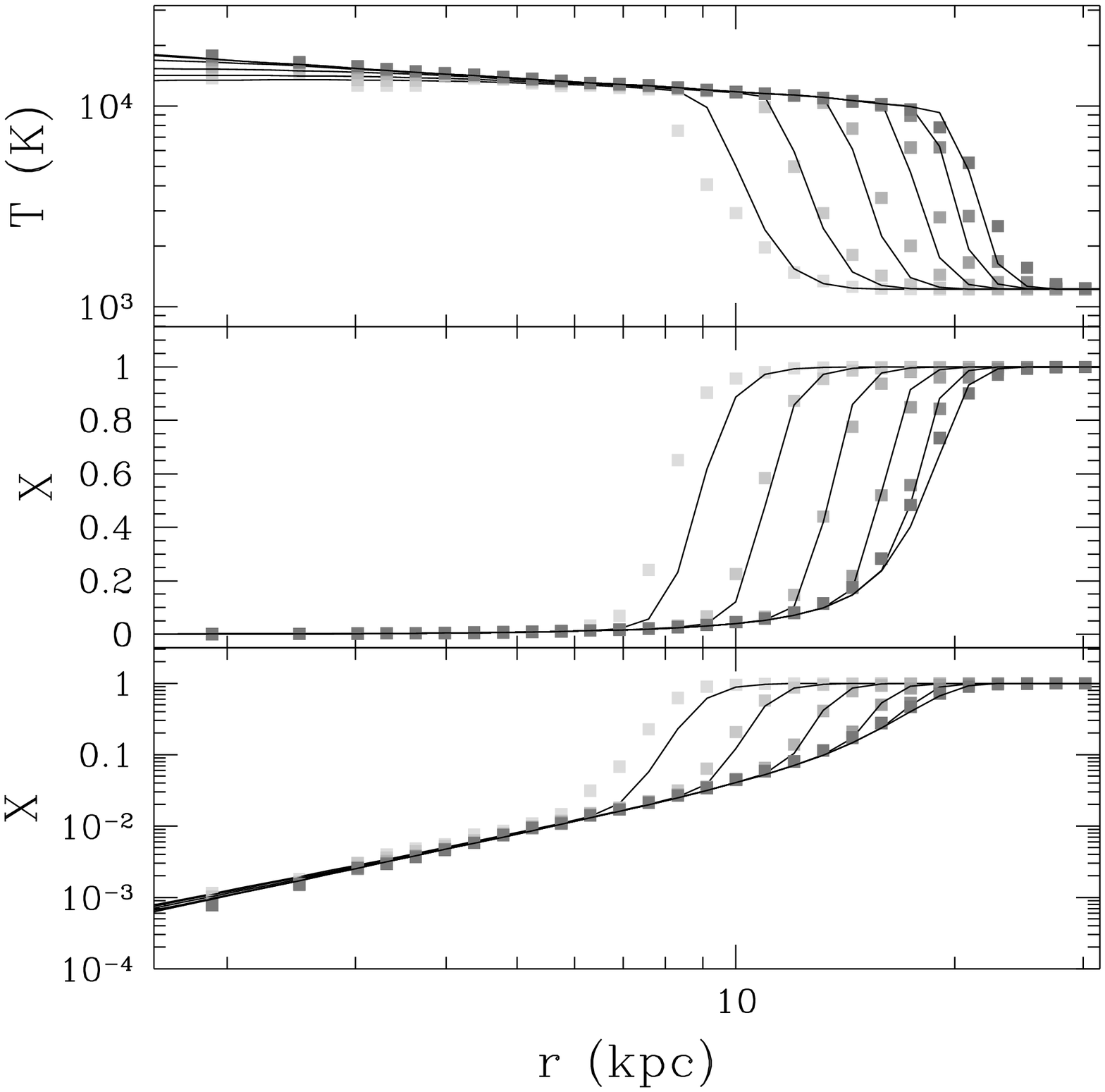}
\caption{\label{figHM}The same as Fig.\ \protect{\ref{figHS}} except
for $\dot{N}_{\rm ph}=10^{50}\dim{s}^{-1}$ ({\it a\/}) and
$\dot{N}_{\rm ph}=10^{48}\dim{s}^{-1}$ ({\it b\/}). In this case
only first 200 hydrodynamic time steps are shown.}
\end{figure}
Figure \ref{figHM} shows in two panels 
the same test with 100 and 10,000 weaker source respectively. Again,
not only the position, but also the detailed profile of the ionization
front is reproduced. In the latter case the source is so weak (by
construction) that it reaches its Stromgen sphere before the ionization
front leaves the box. We continued the weak source test for another
factor of three in time and the solution remains undistinguishable from
the last line shown with no trace of unphysical numerical diffusion
propagating outside of the Stromgen sphere.

\begin{figure}
\includegraphics[width=0.45\hsize]{\figdir/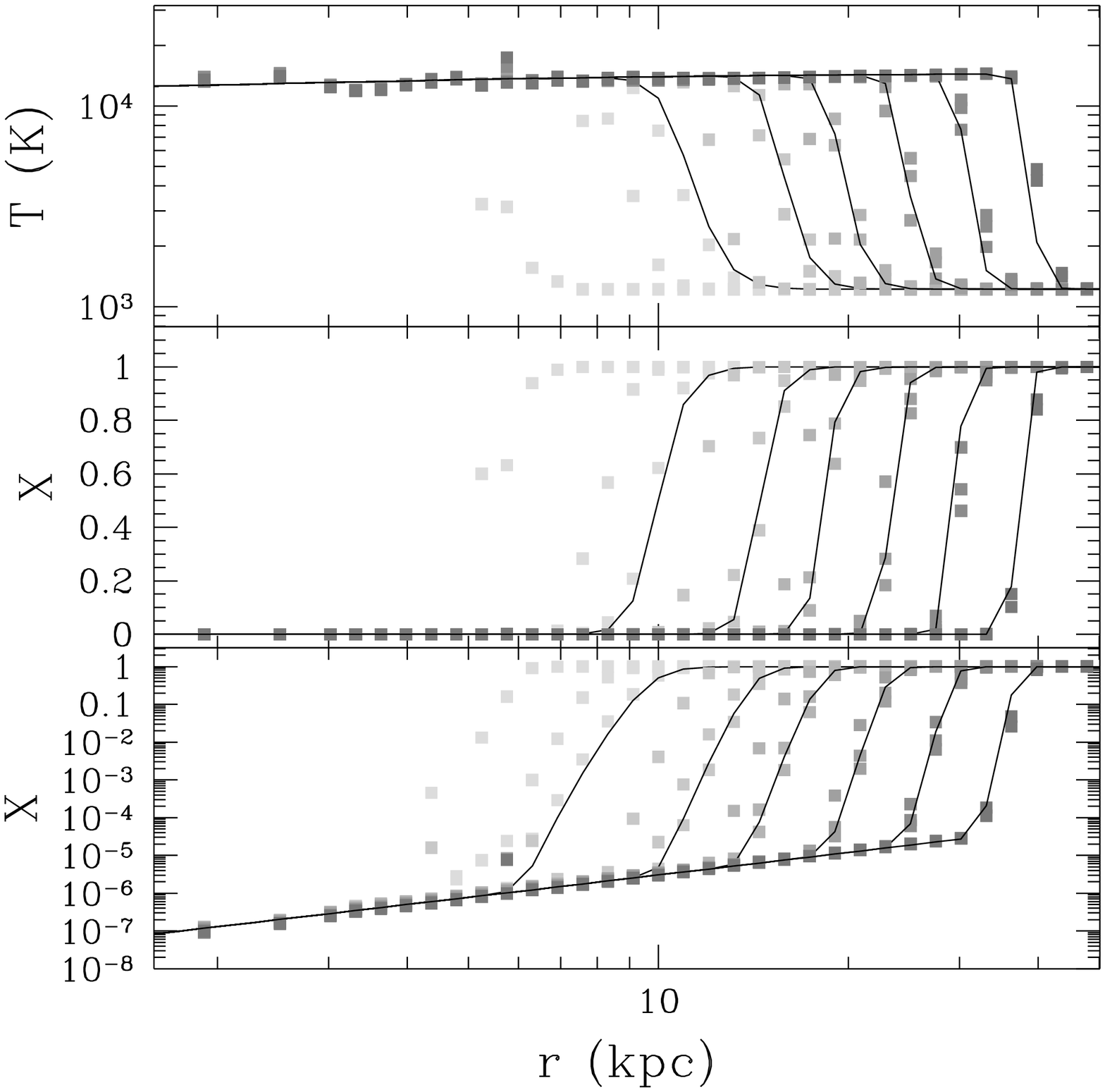}
\includegraphics[width=0.45\hsize]{\figdir/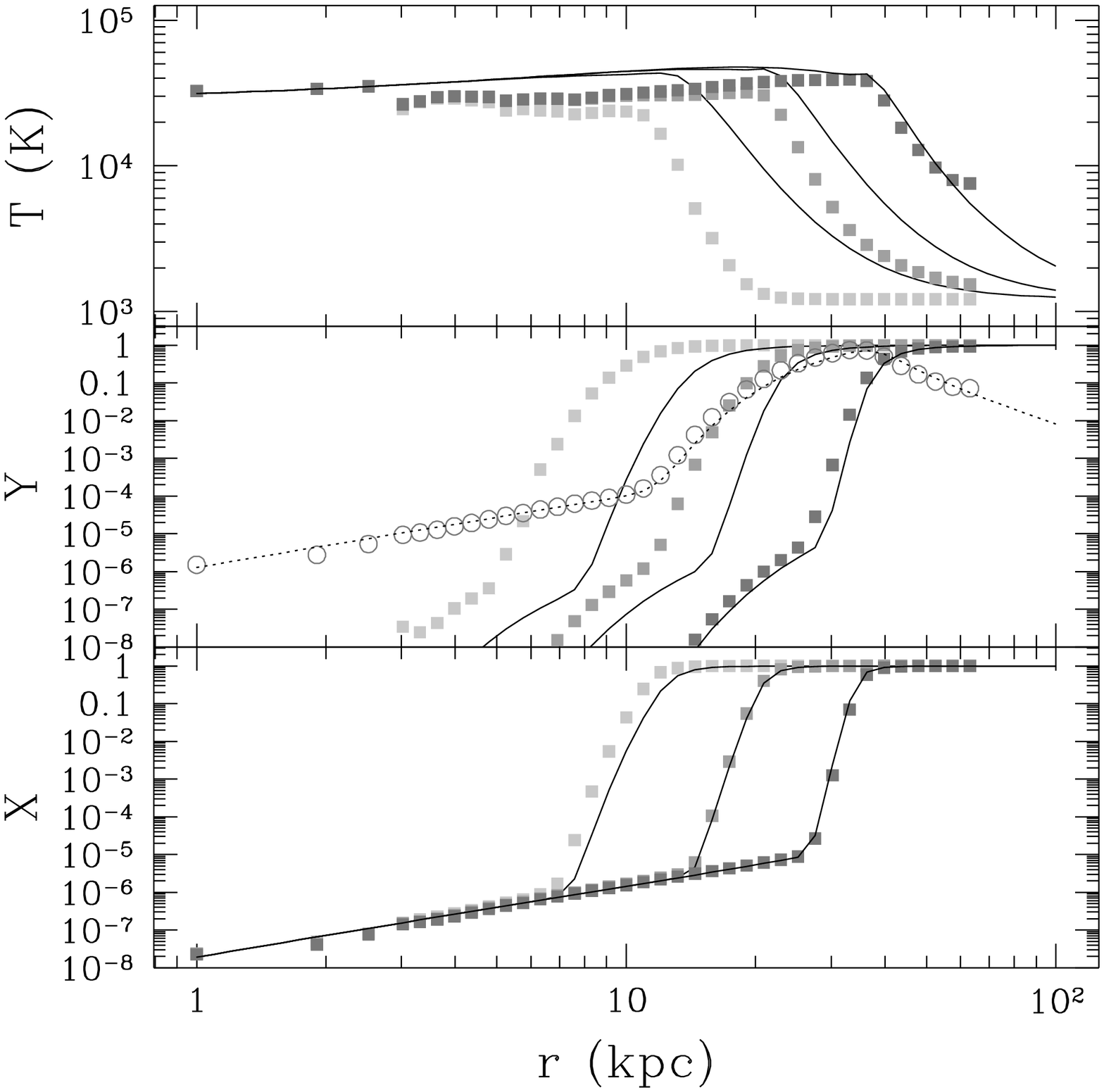}
\caption{\label{figHT}The same as Fig.\ \protect{\ref{figHS}} except
({\it a\/}) three sets of solutions shown corresponding to three
choices for the hydrodynamic time step; ({\it b\/}) the source
spectrum is a power-law with the slope of -1.25 and thus includes
\GI\ and \GII\ ionizing photons. In the latter case the middle panel
shows the \GI\ ionization fraction ({\it filled symbols\/}). Open
symbols mark \GII\ ionization fraction at the last output. Only 6th, 25th,
and 100th hydrodynamic time steps are shown.} 
\end{figure}
Figure \ref{figHT}a shows the effect of changing the hydrodynamic time step.
Three sets of filled symbols correspond to the hydrodynamic time step changed
by a factor of $1/2$, 1, and 2 relative to Fig.\ \ref{figHS}. We again notice
that about 10 hydrodynamic time steps after switching on of the source
are sufficient to achieve complete convergence. Figure \ref{figHT}b also
tests the performance of our scheme in the case when the ionizing spectrum
extends to higher energies, thus being able to ionize \GI\ and \GII\ as well
as \HI. In that case the agreement for helium is somewhat worse than for
hydrogen, and it takes about 30 hydrodynamic time steps to converge on
the helium profiles. This is however is still very small compared to the
total number of hydrodynamic time steps in a realistic simulation, so that
the compound error is expected to be negligible.

\begin{figure}
\includegraphics[width=0.45\hsize]{\figdir/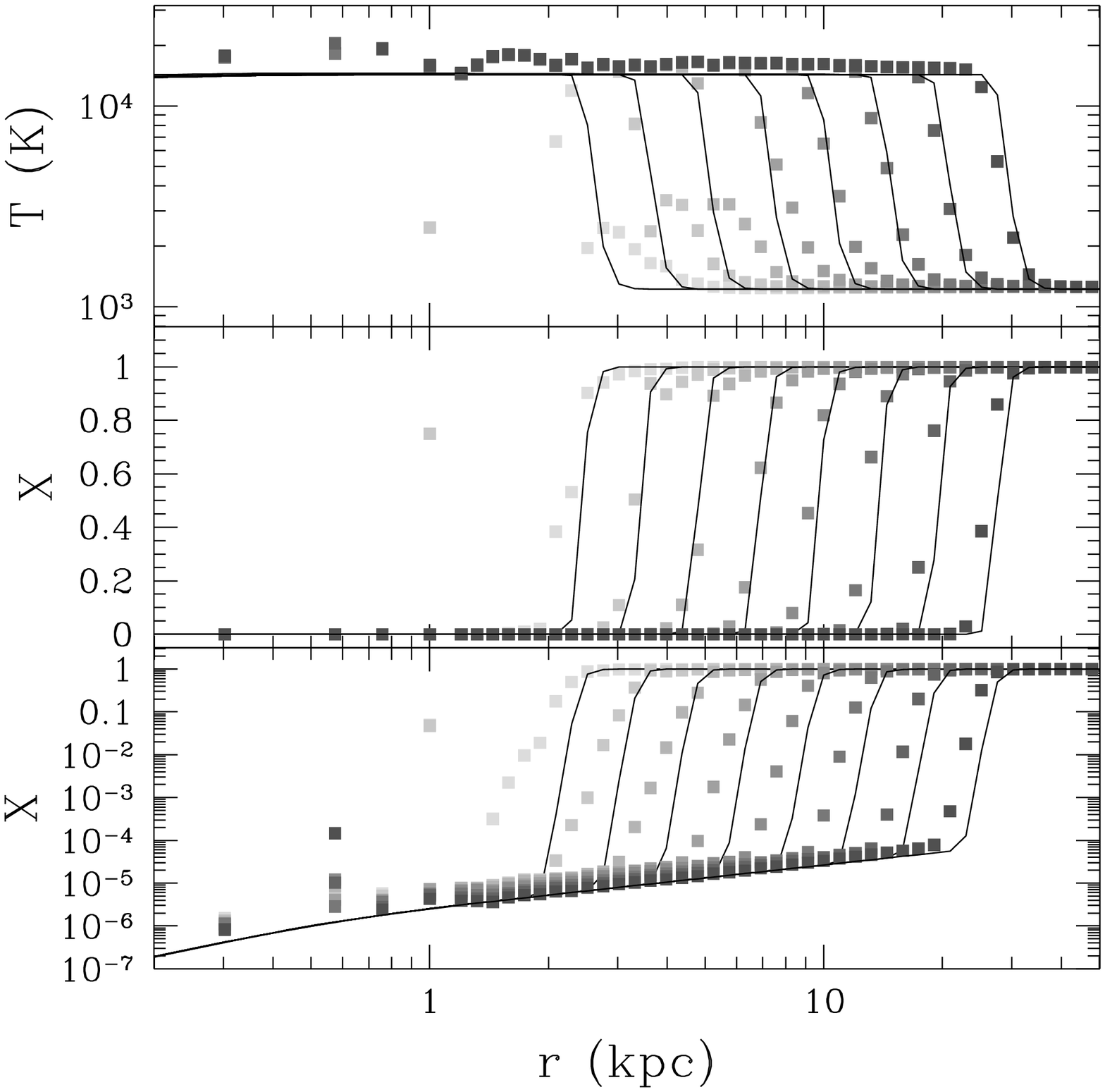}
\includegraphics[width=0.45\hsize]{\figdir/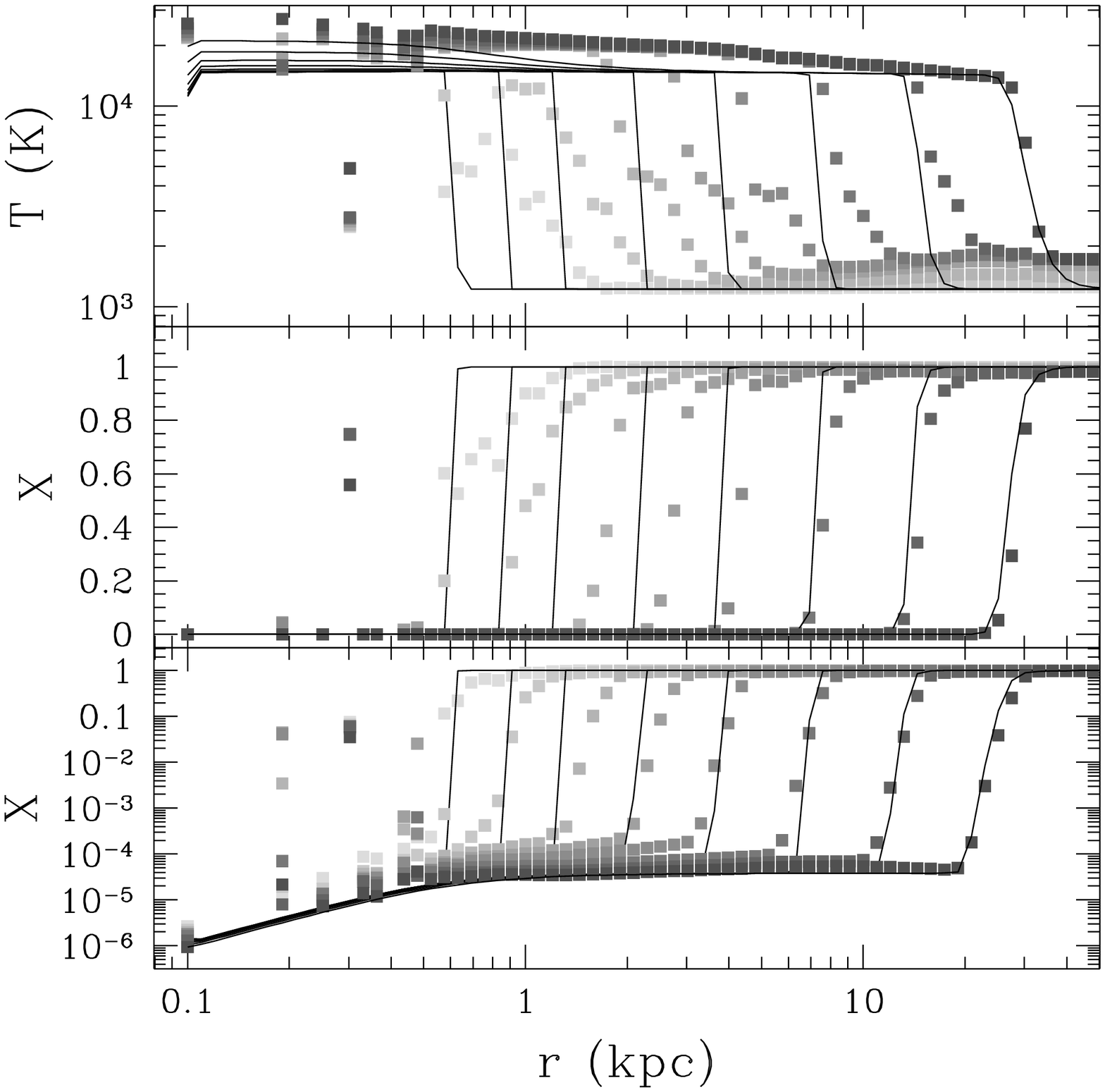}
\caption{\label{figUO}The same as Fig.\ \protect{\ref{figHS}} except
that the density is not uniform by falls off as $r^{-1}$ ({\it a\/}) or
$r^{-2}$ ({\it b\/}) with a small core at the center.}
\end{figure}
Figure \ref{figUO} illustrates another two tests in which the density
distribution is not uniform but falls off away from the source as the
first or the second power of radius. In the latter case the profile of
the ionization front at earlier times is not well reproduced due to the
finite resolution of the 3D scheme.

\subsection{Shadowing}

\begin{figure}
\includegraphics[width=\hsize]{\figdir/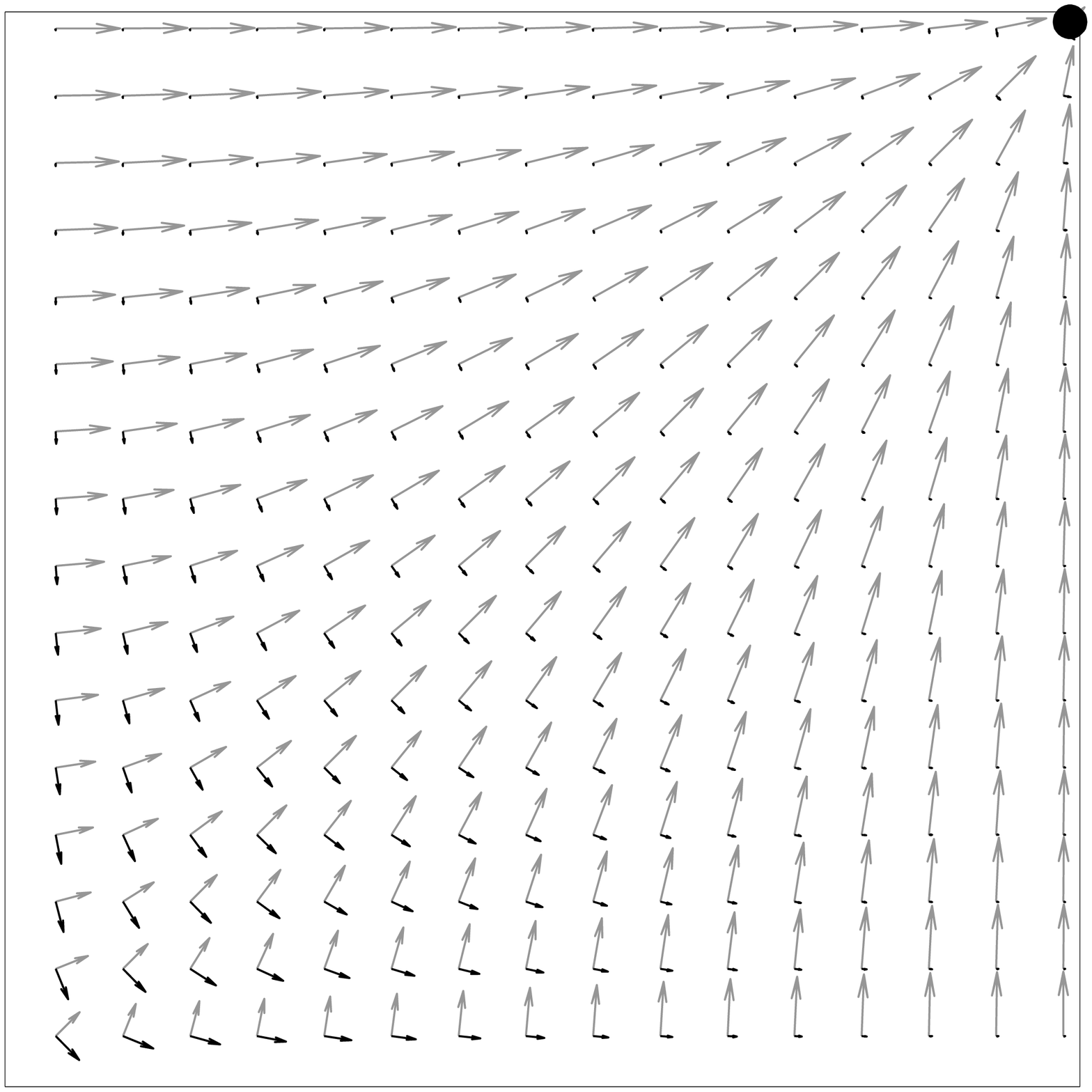}
\caption{\label{figTN}Two eigenvectors of the optically thin Eddington
tensor in one quadrant of $z=0$ section of the computational box.
The source location is labeled by a filled circle.}
\end{figure}

The tests from the previous section demonstrate that our scheme performs
reasonably well for a single source in a spherically symmetric case.
However, the choice of the optically thin Eddington tensor is, obviously,
an approximation, and we can identify regimes where it is likely to fail.
This is best illustrated by Figure \ref{figTN}, which shows the two
eigenvectors of the optically thin Eddington tensor for a single source.
The box has a unit side, and is located so that its center coincides with
the point (0,0,0). The source is located at the center. We show the $x-y$
plane of the box with two eigenvectors laying in that plane. The eigenvectors
are scaled by the square root of their respective eigenvalues  
(due to symmetry,
the third eigenvector is directed along $z$ direction, but the respective
eigenvalue is zero). Their signs therefore are undetermined, i.e. each vector
can be turned around 180 degrees.

In this test case it is easy to find the exact Eddington tensor - it 
has only one nonvanishing eigenvalue (therefore equal unity everywhere
because the trace of the Eddington tensor is identically equal unity)
and the corresponding eigenvector at
every point in space points towards (or away from) 
the source (i.e.\ $h^{ij}=e^i_re^j_r$). The grey vectors in 
Fig.\ \ref{figTN} show this eigenvector. However, in the optically
thin case there exist another eigenvector, shown in black in 
Fig.\ \ref{figTN}. This eigenvector arises due to periodic boundary
conditions as a contribution from
periodic images of the source. As the result, the second eigenvector
will allow for the propagation of the ionization front in the direction
tangential to the source (albeit with a lower speed).

\begin{figure}
\includegraphics[width=\hsize]{\figdir/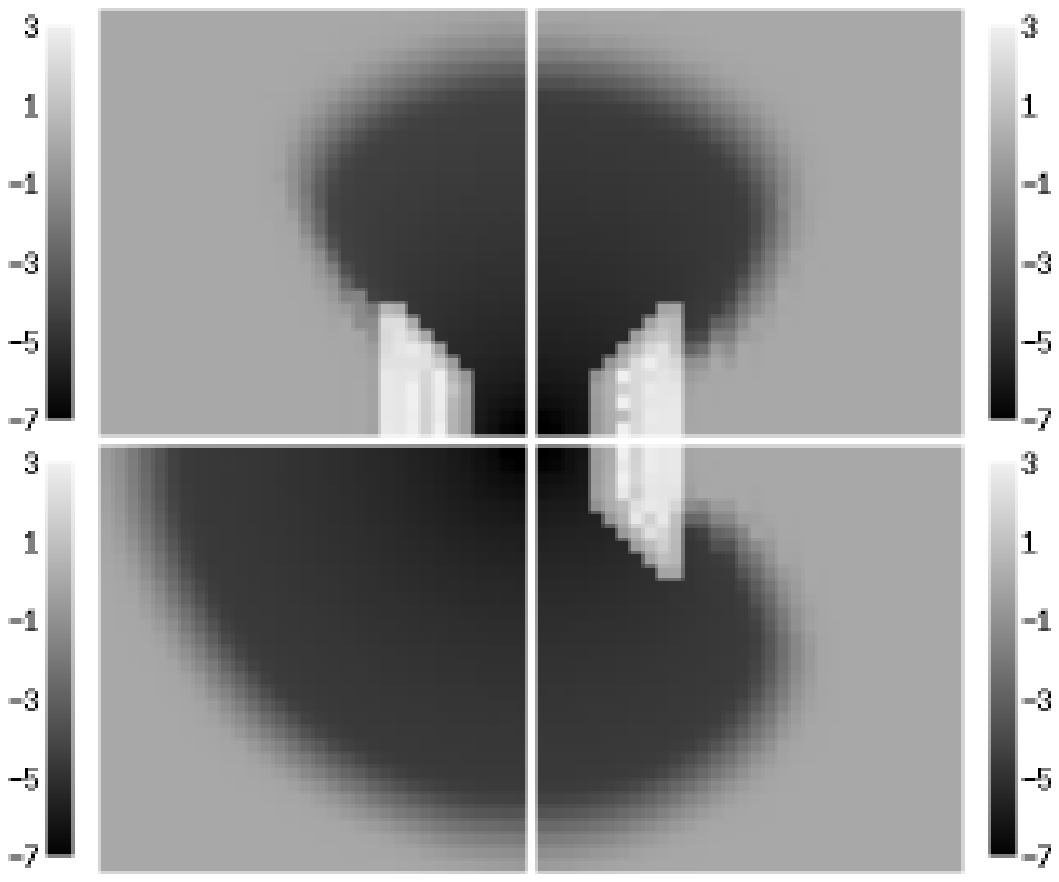}
\caption{\label{figWE} The distribution of neutral hydrogen in the one
quadrant of the midplane of the computational box for four different tests:
uniform medium ({\it lower left panel\/}),
uniform medium with a 1000 times denser wedge ({\it lower right panel\/}),
uniform medium with a wedge in a twice larger ($128^3$) box 
({\it upper right panel\/}),
uniform medium with a wedge with the exact Eddington tensor
({\it upper left panel\/}). All panels are flipped so that the source is
located at the center of the plot. The side bars give the correspondence
between the color and the decimal log of the neutral hydrogen density.}
\end{figure}
Figure \ref{figWE} serves to illustrate this point. Its lower left panel
shows the one quadrant in the midplane from a test shown in Fig.\ 
\ref{figHS}.
Dark color shows the ionized region, and light color shows the still neutral
medium. The lower right panel shows the same test (the panel is flipped
around the $y$ axis) but with the 1000 times denser wedge inserted on the
way of the ionization front. The wedge is supposed to shield the medium
behind it from the ionizing radiation, however we can see the ionization
front propagating behind the wedge with about 1/5 of the normal ionization
front speed. This propagation is caused by the second eigenvector of 
the Eddington tensor, and is therefore unphysical - it is a manifestation of
the optically thin Eddington tensor calculated in a periodic universe.

To verify that our conclusion is indeed correct, we have performed the same
test in a twice larger box ($128^3$ instead of $64^3$) shown on the upper 
left panel of Fig.\ \ref{figWE}. In that case all
the periodic images are twice further away, the contribution from them to
the Eddington tensor is 4 times smaller, and we expect the tangential
ionization front to propagate with $(1/4)^{1/3}=0.6$ times lower speed,
as indeed can be observed from the figure. 

Finally, we show the same test but calculated with the exact Eddington
tensor on the upper left panel. In this case the ionization front does
not propagate behind the wedge except by a very weak numerical diffusion
(which is not important in a realistic simulation).

These tests illustrate the main shortcoming of the proposed scheme:
the slow propagation of the ionization fronts behind the shadow. We
emphasize that this propagation is quite slow, significantly slower
than the ionization front speed. We notice this effect only because
there should be no propagation of the radiation behind the shadow
whatsoever. Our tests are not sufficient to evaluate whether this is a
significant drawback or it is not important in realistic simulations -
we expect that future work will shed light on this question.

In the proposed framework, we are not aware of the acceptable way to
circumvent this deficiency of the scheme. We have tried to experiment
with different choices of the Green functions for calculating the
Eddington tensor, but found no noticeable improvement - with the
following simple reason: if we adopt a Green function which falls
faster than $1/r^2$ (a short-range one) in order to reduce the
contributions from the distant periodic images, we at same time
introduce deformations into the Eddington tensor inside the
computational box, because of the periodicity of boundary
conditions. These deformations may lead to even larger errors than the
original optically thin Eddington tensor. Significant improvement over
the proposed scheme is only possible if the Eddington tensor is
estimated more accurately, for example, by calculating the exact
tensor at a number of selected points inside the computational
box. However, it seems unlikely this could be done with acceptable
computational cost.


We again emphasize here that the fact that the Eddington
tensor is calculated in the optically thin regime does not compromise
shadowing by itself, but only due to imposed periodic boundary
conditions.  The reason that the shadowing works with the optically
thin Eddington tensor is because the absorption of ionizing photons by
the obstacle is computed essentially exactly (up to the RSL
approximation, which becomes exact in the limit of a very dense
obstacle).  The optically thin approximation does make an error by
assigning the incorrect values for the Eddington tensor behind the
obstacle, but because the moments equations insure that the ionizing
photon number density behind the obstacle is zero, the error in the
Eddington tensor becomes irrelevant.

\subsection{Multiple sources}

It may appear that the choice of the optically thin Eddington tensor
will also lead to a serious error in a case of several sources inside
the computational box. Indeed, let's consider two sources inside a
computational box: a strong one and a weak one, and let's assume that
the two sources switch on simultaneously. Until their respective
ionization fronts meet, each will have a spherical \HII\ region around
it, and inside each of the \HII\ regions the Eddington tensor will be
of the form $n^in^j$ where $\vec{n}$ is the unit vector pointing
toward the respective source.  However, the proposed scheme adopts an
optically thin Eddington tensor, which means that at each point in
space the Eddington tensor is given by equation (\ref{ouret}) with
\[
	P^{ij} = 
	{S_1 r_1^i r_1^j\over r_1^4} + 
	{S_2 r_2^i r_2^j\over r_2^4}, 
\]
where $\vec{r}_1$ and $\vec{r}_2$ are coordinates of the two sources in a
reference frame tied to a given point in space, and $S_1$ and $S_2$ are
their respective luminosities. We assume that $S_1>S_2$.

Near the strong source the contribution from the weak source is small,
and thus the optically thin Eddington tensor is a good approximation
to the exact one. However, near the weak (second) source the
contribution from the first source could be important: even if
$r_2<r_1$, $S_1>S_2$. However, the physics of the ionization fronts is
such that this is not a significant problem. We can illustrate that
with the following simple estimate.

The radius of the ionization front scales as a cubic root of the
source luminosity. Thus, at the moment when the ionization fronts from
the two sources overlap, the ratio of the front radii is
\[
	R1/R2 = (S1/S2)^{1/3}.
\]
The contribution to the tensors from each of the sources is proportional to 
$S/r^2$, i.e.\ the ratio of the contributions from the weak and the 
strong sources to the Eddington tensor within the weak source's 
\HII\ region is
\[
	{h^{ij}_2\over h^{ij}_1} = {S_2 r_1^2\over S_1 r_2^2}.
\]
\begin{figure}
\includegraphics[width=\hsize]{\figdir/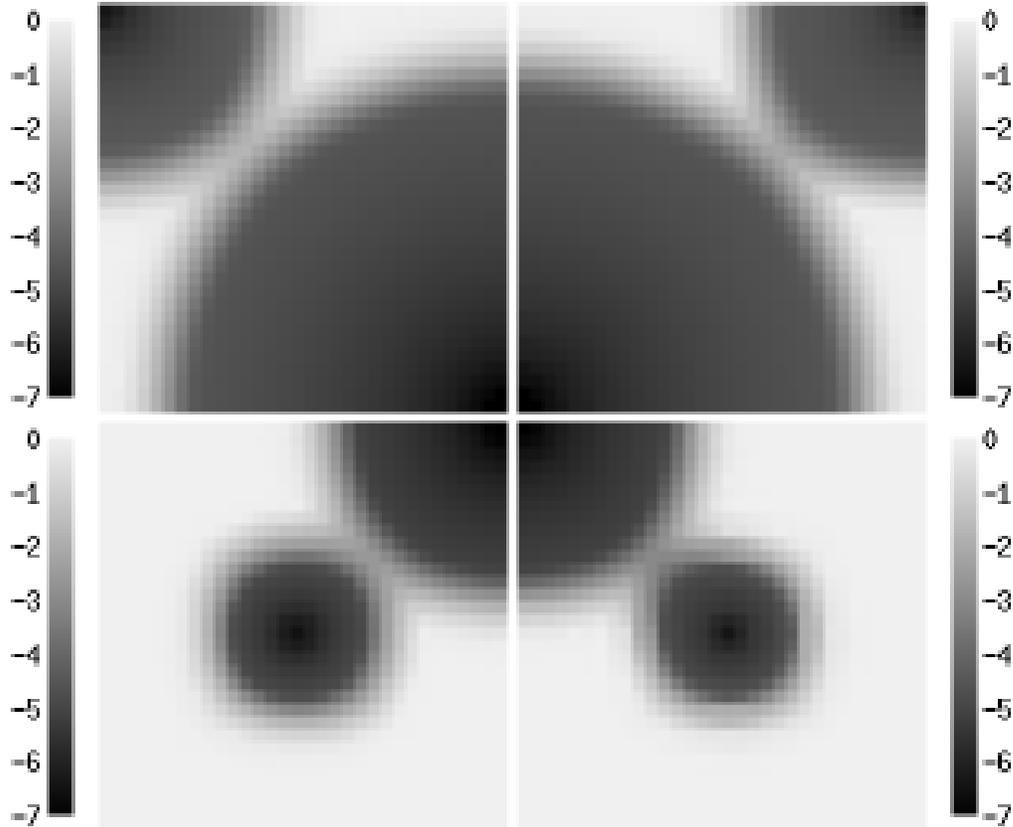}
\caption{\label{figTS} The distribution of neutral hydrogen in one
quadrant of the midplane of the computational box for four different cases.
The upper row shows the two source test with the weak source located at
(-0.5,-0.5,0) and in the bottom row the weak sources is at (-0.25,-0.25,0).
The strong source is at (0,0,0), and the ratio of source luminosities is 10.
Left columns show the case with the exact Eddington tensors (which can only
be calculated easily before the two ionization fronts overlap), and right
columns show the test with optically thin Eddington tensors. In both cases
the difference is noticeable but not large.}
\end{figure}
This contribution is greater 
than 1 (i.e. the weak source dominates its own Eddington 
tensor in the optically thin regime regime) when
\[
	\left(r_2\over R_2\right)^2 < \left(r_1\over R_1\right)^2
	{R_2\over R_1}
\]
or
\[
	{r_2\over R_2} < {r_1\over R_1}  \left(S_2\over S_1\right)^{1/6}.
\]
Even at the time of overlap, when $r_1\sim R_1$, 
most of the volume inside the HII 
region around the weak source is dominated by the weak source, because $1/6$ 
is too weak a power law. Before the overlap, $r_1>R_1$ ($r_1$ is 
the distance from the strong source to a point inside the weak source 
\HII\ region), so the constraint is even weaker.

\begin{figure}
\includegraphics[width=\hsize]{\figdir/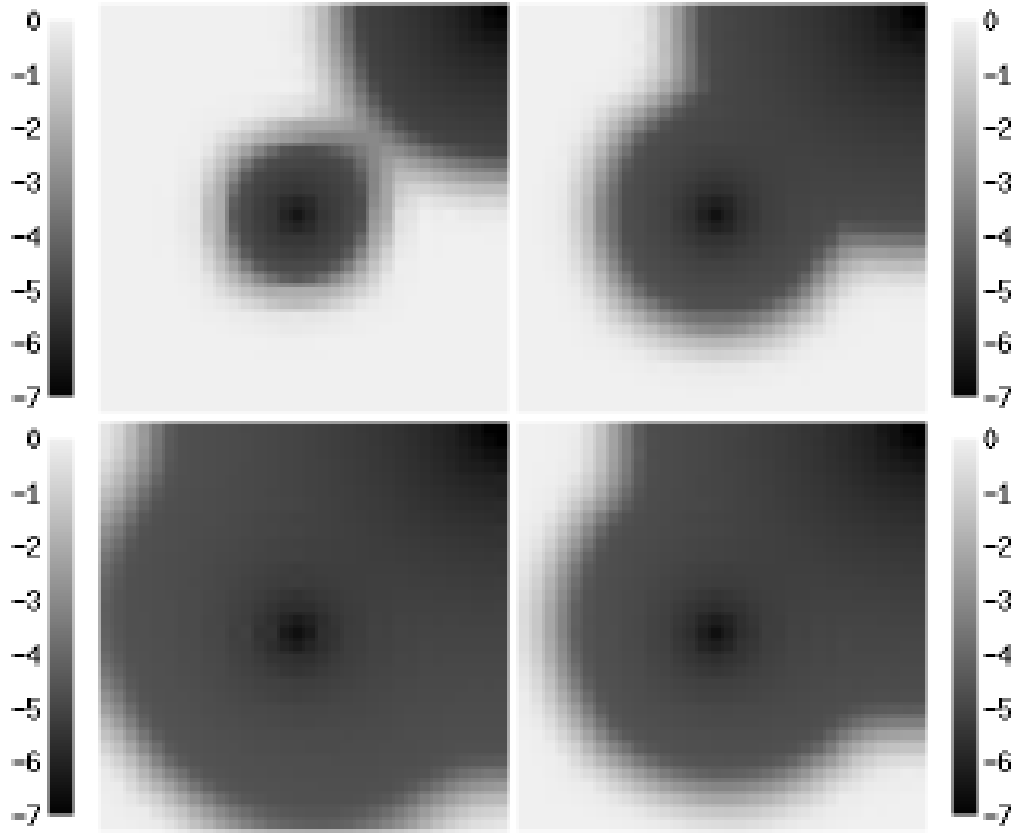}
\caption{\label{figTT} The distribution of neutral hydrogen in one
quadrant of the midplane of the computational box for 
the two-source test with optically thin Eddington tensors at four
different times.}
\end{figure}
Figure \ref{figTS} illustrates this conclusion. The two two-source tests
presented in the figure show that the effect of the stronger source on the 
propagation of the ionization front around a weaker source are small
albeit still noticeable. We also show in Figure \ref{figTT} consequent
evolution for this test for the approximate solution. We are unable to
compute the exact solution in this case because we cannot calculate the
exact Eddington tensor when two \HII\ regions overlap.

\subsection{Optically thin regime}

\begin{figure}
\includegraphics[width=\hsize]{\figdir/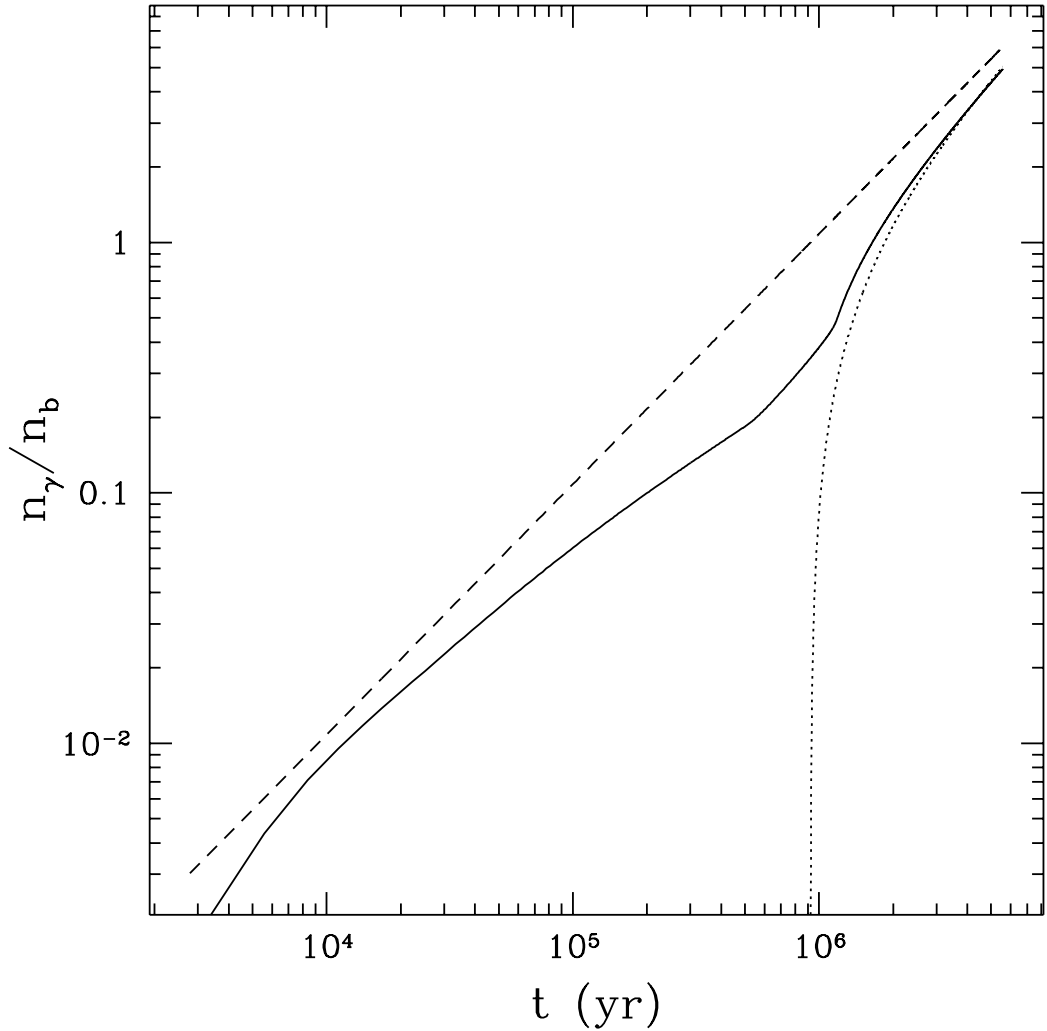}
\caption{\label{figNB} The number of ionizing photons per baryon 
as a function of time for the optically thin case
({\it dashed line\/}) and for the full calculation 
({\it solid line\/}) for the test presented in Fig.\ \protect{\ref{figHS}}.
The dotted line shows the optically thin case minus one (one photon
per baryon is used to ionize the box).}
\end{figure}
One of the attractive features of this scheme is that the transition
to the optically thin regime (or, more precisely, to the regime when
the mean free path is much larger than the size of the computational
box) is automatic, whereas in many alternative approaches it is not so
trivial.  Fig.\ \ref{figHS} shows this regime for one of our tests
with open symbols.  In Figure \ref{figNB} we show the ionizing
background (i.e.\ the volume averaged specific intensity) measured in
units of ionizing photons per baryon in the computational box for the
optically thin case (no absorptions) and for one of our single source
tests. The dotted line in this figure shows the optically thin case
minus one, which approaches the full solution at later times. This
illustrates the conservation of photons as the solution switches from
the optically thick to the optically thin regime (in this test,
recombinations are not important, and only one ionizing photon per
baryon is needed to ionize the whole box).

This test illustrates that our scheme automatically insures the exact 
conservation of the number of photons as the mean free path crosses the
box boundaries.

\subsection{Realistic simulations}
\label{sim}

\begin{figure}
\includegraphics[width=\hsize]{\figdir/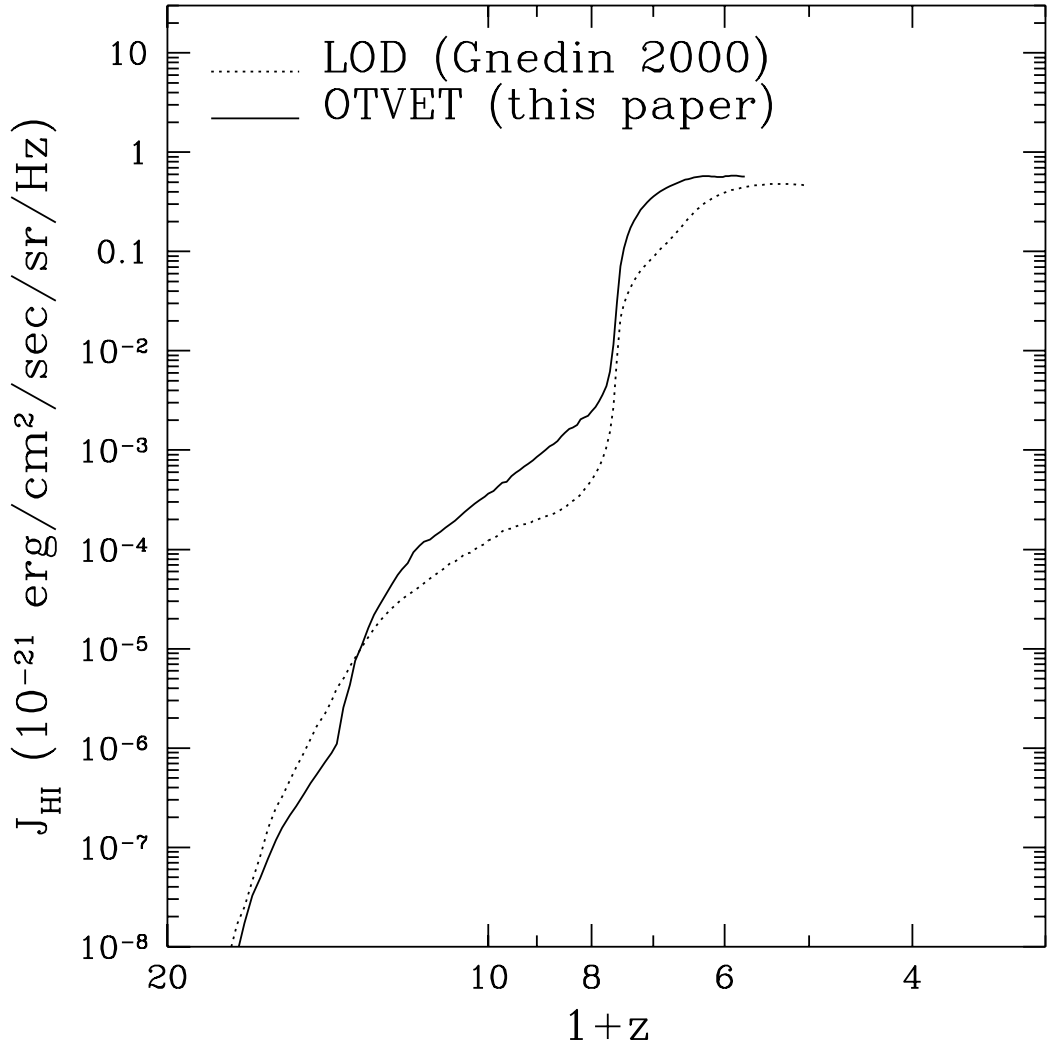}
\caption{\label{figJJ} 
The evolution of the spatially averaged ionizing intensity 
$J_{\mbox{\small\HI}}$
(measured in conventional units of 
$10^{-21}\dim{erg}/\dim{cm}^2/\dim{sec}/\dim{Hz}/\dim{rad}$) 
in two simulations of a representative cosmological model: using
the Local Optical Depth (LOD) approximation of 
\citet[][{\it dotted line\/}]{G00}
 and Optically Thin Variable Eddington Tensor
(OTVET; {\it solid line\/}) approximation developed in this paper.}
\end{figure}
While the main purpose of this paper is to present the new and
accurate approximation for modeling continuum radiative transfer, we
give here also an example of how it performs for the problem it was
developed for, cosmological reionization. For this purpose we have
repeated one of the simulations from \citet{G00}, namely the run
labeled ``N64\_L2\_A''.  That simulation contained $64^3$ dark matter
particles, an equal number of baryonic cells and, towards the end of
the calculation, roughly the same number of stellar particles in a
computational box with the size of $2h^{-1}$ comoving Mpc.  A reader
can find a full description of this simulation in \citet{G00} and we
do not repeat it here for brevity. This test allows us not only to
check the performance of our Optically Think Variable Eddington Tensor
(OTVET) approximation in a realistic cosmological simulation, but also
to compare it with the much less accurate Local Optical Depth (LOD)
approximation. The OTVET radiative transfer simulation presented here
consumed roughly three times the computational time than the one using
LOD transfer.

In Figure \ref{figJJ} we show the evolution of the volume averaged
ionizing intensity for two simulations. The two simulations have
identical parameters and the same physical content - the only
difference is in the method used to solve for the radiative transfer.
We notice here that the two simulations are qualitatively similar, and
reionization (the sharp rise in the specific intensity at around
$z=7$) occurs almost at the same time in both simulations. This should
be expected because the LOD approximation is designed to give the
correct time for the overlap of \HII\ regions, which marks the time of
reionization. However, there are also noticeable differences, which we
attribute to the inaccuracy of the LOD approximation.

\begin{figure}
\includegraphics[width=\hsize]{\figdir/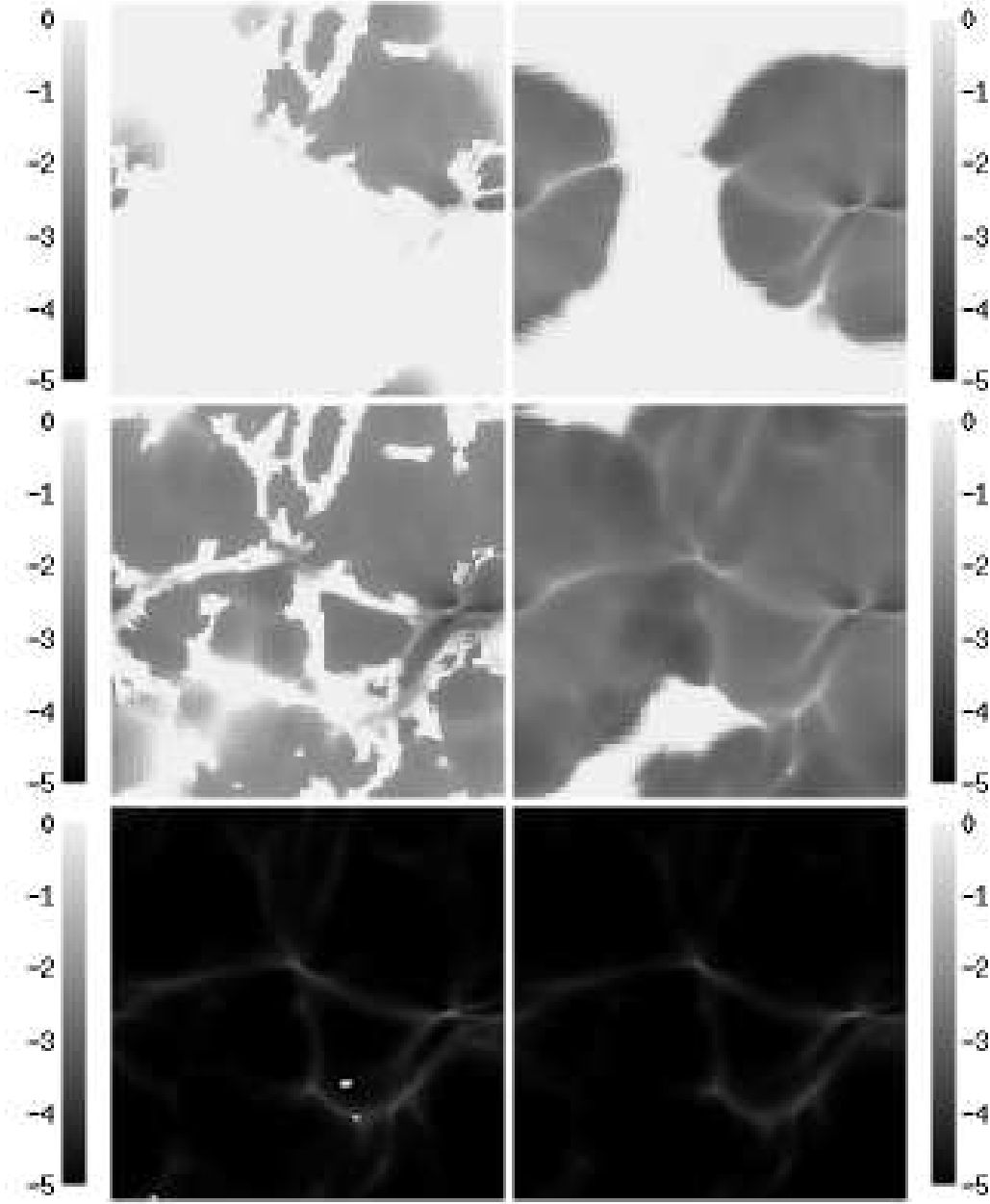}
\caption{\label{figRE} 
The distribution of neutral hydrogen in a realistic simulation of
cosmological reionization at three different times as shown in
three rows: before the overlap of \HII\ regions ({\it top row\/}),
during the overlap ({\it middle row\/}), and after the overlap
({\it bottom row\/}). The left column shows the highly approximate
solution with the Local Optical Depth (LOD) approximation from
\citet{G00}, and the right column shows the present scheme (OTVET).}
\end{figure}
In order to investigate these differences further, we show in Figure
\ref{figRE} slices from two simulations at three different times -
before the overlap of \HII\ regions, during the overlap, and after the
overlap.  We notice that while large-scale features are similar in two
simulations, small-scale structure is quite different. The main
difference is the excess of low density neutral gas in the LOD
approximation. This is indeed expected, because the LOD approximation
makes the largest error in the regions with the optical depth is about
unity.

It appears from Fig.~\ref{figRE} that this error is systematic -
neutral regions are always larger in the LOD approximation - and thus
the LOD approximation performs worse than one of the authors (NG)
hoped for. On the other hand, the LOD approximation only reproduces
the ionization fronts speed on average, i.e.\ it does not guarantee a
correct speed for every ionization front. The fact that the
large-scale structure of ionization fronts agrees quite well between
the two simulations indicate that the ionization front speed in the
LOD approximation is calculated more accurately than could be expected
a priori. This observation would allow to improve the LOD
approximation by correcting the systematic deviation on the
$\tau\sim1$ regime. 

Another significant difference between the two approximations is a
higher asymmetry of \HII\ regions at early times in the LOD
approximation.  This difference arises because the LOD approximation
estimates the total optical depth to the source from its local
density. Somewhat denser material becomes therefore artificially more
difficult to ionize. Consequently, I--fronts wrap around dense
material unphysically strongly giving rise to the strongly distorted
ionization fronts observed in Figure~\ref{figRE}.


Finally, at later times (after the overlap) the two solutions are
quite similar (except for small errors in the LOD approximation in the
$\tau\sim1$ regime) because both approximations become exact in the
optically thin regime.

\section{Conclusions} 

We present a new approach to numerically model continuum radiative
transfer based on the Optically Thin Variable Eddington Tensor (OTVET)
approximation. Our method insures the exact conservation of the photon
number and flux (in the explicit formulation) and automatically
switches from the optically thick to the optically thin regime. It
scales as $N\log N$ with the number of hydrodynamic resolution
elements and is independent of the number of sources of ionizing
radiation (i.e.\ works equally fast for an arbitrary source function).

We also describe an implementation of the algorithm for the Soften
Lagrangian Hydrodynamic code (SLH). We present extensive tests of our
method for single and multiple sources in homogeneous and
inhomogeneous density distributions, as well as a realistic simulation
of cosmological reionization.

The method fully solves the cosmological radiative transfer equation
in a conservative fashion via moment equations. The only approximation
is our use of optically thin Eddington tensors. One may envision to
employ a Monte Carlo or short characteristic methods to supply more
accurate tensors for applications in which an even more accurate
knowledge of the isotropy of radiation field is required.

The main shortcoming introduced by optically thin Eddington tensors
with periodic boundary conditions is the unphysical propagation of
ionization fronts behind an optically thick obstacle. This diffusive
effect is however slow and negligible for the typical R--type
ionization fronts of interest in numerical cosmology.

A realistic simulation of cosmological reionization with our current
implementation of the new method took approximately three times the
computing time than a simulation without radiative transfer. However,
there are several optimizations which we have not yet implemented,
which will shorten this time by at least a factor of two.

T.A.\ acknowledges stimulating and insightful discussions with Markus
Ramp, Dimitri Mihalas, Pascal Paschos and Michael Norman. 
We are also grateful to the referee Andreas Burkert for friendly and
valuable comments. This work
has partially been supported through NSF grants ACI96-19019 and
AST-9803137, as a part of GC$^3$, and by National Computational
Science Alliance under grant AST-960015N, and utilized the SGI/CRAY
Origin 2000 array at the National Center for Supercomputing
Applications (NCSA).

\end{document}